\documentclass[12pt]{article}
\input feynman.tex
\usepackage{epsfig}
\usepackage{amssymb}

\newcommand{\f}{\frac}
\newcommand{\la}{\lambda}
\newcommand{\rla}{\sqrt{\lambda}}
\newcommand{\be}{\begin{equation}}
\newcommand{\beq}{\begin{equation}}
\newcommand{\ee}{\end{equation}}
\newcommand{\eq}{\end{equation}}
\newcommand{\bea}{\begin{eqnarray}}
\newcommand{\eea}{\end{eqnarray}}
\newcommand{\rb}{\underline{r}}
\newcommand{\kb}{\underline{k}}
\newcommand{\lb}{\underline{l}}

\newcommand{\ab}{\underline{a}}
\newcommand{\bb}{\underline{b}}
\newcommand{\pb}{\underline{p}}

\newcommand{\Kb}{\underline{K}}
\newcommand{\Rb}{\underline{R}}
\newcommand{\zb}{\bar{z}}

\newcommand{\aab}{\bar{a}}
\newcommand{\lr}{\leftrightarrow}
\newcommand{\eps}{\epsilon}
\newcommand{\intfeyn}{\int\limits_0^1}

\begin{document}

\begin{flushright}
 CPHT-RR 042.0605 \\
LPT 04-117 \\
  hep-ph/0507038
\end{flushright}

\vspace{\baselineskip}

\begin{center}
\textbf{\LARGE Double diffractive $\rho-$production in $\gamma^*\gamma^*$
  collisions 
%1:\\
% Born order
}\\

\vspace{3\baselineskip}

\vspace{3\baselineskip}
{\large
B. Pire$^a$,
 L. Szymanowski$^{b,c}$ and S. Wallon$^d$
}
\\

\vspace{2\baselineskip}
${}^a$\,CPHT\footnote{Unit{\'e} mixte C7644 du CNRS}, \'Ecole
Polytechnique, 91128 Palaiseau, France \\[0.5\baselineskip] 
${}^b$\,Soltan Institute for Nuclear Studies, Warsaw, Poland
\\[0.5\baselineskip] 
${}^c$\,Universit\'e  de Li\`ege,  B4000  Li\`ege, Belgium\\[0.5\baselineskip]
${}^d$\,LPT\footnote{Unit{\'e} mixte 8627 du CNRS}, Universit\'e Paris-Sud, 91405-Orsay, France  \\

\vspace{5\baselineskip}
\textbf{Abstract}\\
\vspace{1\baselineskip}
\parbox{0.9\textwidth}{We present a first estimate 
of the cross-section for the exclusive process 
$\gamma^{ *}_L (Q_1^2)\gamma^{ *}_L(Q_2^2) \to
\rho_L^0 \rho_L^0,$
which will be studied in the future high energy  $e^+e^--$ linear collider. As
a first step, 
we calculate the Born order approximation of the amplitude for longitudinally
polarized virtual photons and mesons, in the kinematical region $s \gg -t,
\,Q_1^2 \,, Q_2^2.$
This process is completely calculable in the hard region $Q_1^2 \,, Q_2^2 \gg
\Lambda^2_{QCD}.$ We perform most of the steps in an analytical way.
The resulting cross-section turns out to be large enough for this process to
be measurable with foreseen luminosity and energy, for $Q_1^2$ and $Q_2^2$ in the range of a
few ${\rm GeV^2}.$ 

}

\end{center}

\medskip
\noindent
\section{ Introduction}
\setcounter{equation}{0}

The next generation of $e^+e^--$colliders will offer a possibility of
clean testing of QCD dynamics. 
 By selecting events in which two vector mesons are produced
with large rapidity gap, through scattering of two highly virtual photons,
one is getting access to the kinematical regime in which the perturbative
approach is justified. If additionally one selects the events with
comparable photon virtualities, the perturbative Regge dynamics of QCD
 of the BFKL \cite{bfkl} type, should dominate with respect to the
conventional partonic evolution of DGLAP \cite{dglap} type.
Apart from the study of
the total cross section which has been proposed as a test of BFKL dynamics
 \cite{bfklinc,royon},
one can achieve similar goals by studying diffractive reactions.
From this point of view the production of two $J/\Psi-$mesons was studied
in Ref. \cite{2jpsi}. In this case the hard scale is supplied by the mass of the
heavy quark. In the present paper we propose to study the electroproduction
of two  $\rho-$mesons in the $\gamma^* \gamma^*$
collisions. 
In this case the virtualities of the scattered photons play the role of the hard 
scales. 
As a first step in this direction we shall consider this
process with  longitudinally polarized photons and 
$\rho-$mesons,
 \beq
\label{process}
\gamma^*_L(q_1)\;\gamma^*_L(q_2) \to \rho_L(k_1)  \;\rho_L(k_2)\,,
\ee
for arbitrary values of $t=(q_1-k_1)^2,$ with $s \gg -t.$ 
 The choice of longitudinal polarizations of both the
scattered photons and produced vector mesons is dictated by the
fact that this configuration of the lowest twist-2 gives the dominant
contribution in the powers of the hard scale $Q^2,$ when $Q_1^2 \sim Q_2^2
\sim Q^2$. 
The measurable cross section is related to the amplitude of this process
 through the usual photon flux factors :
 \beq
\label{eesigma}
 \frac{d\sigma(e^+ e^- \to e^+ e^- \rho \rho)}{dy_{1}dy_{2}dQ_{1}^2dQ_{2}^2}
= \frac{1}{Q_1^2Q_{2}^2}\frac{\alpha}{2\pi} P_{\gamma/e}(y_{1}) P_{\gamma/e}(y_{2}) 
\sigma(\gamma^*\gamma^*\to \rho \rho)\;,
\ee
where $y_{i}$ are the longitudinal momentum fractions of the bremsstrahlung photons with 
respect to the respective leptons and with $P_{\gamma/e}(y) = 2(1-y)/y$¥ for longitudinally 
polarized photons. These double tagged events allow to access this hard
cross-section \cite{budnev}.
According to our best knowledge this process has not been discussed up to
now. 
At lower energy, some experimental data exist for $Q_2^2$ small  \cite{L3} and
may be 
analysed \cite{apt} in terms of generalized distribution amplitudes
\cite{gda} or transition distribution amplitudes \cite{tda}.

In this paper we calculate
  the scattering amplitude of the process (\ref{process}) in the Born
approximation. In this way we get an estimate of the 
cross section and prove the feasibility of a dedicated experiment. Partial results have been presented in Ref.\cite{conf}.  
In the near future, we intend to extend this study by taking into account of BFKL
  evolution and transverse photon polarizations.

\section{Kinematics}
\label{kinematics}
\setcounter{equation}{0}

\begin{figure}[htp]
\begin{picture}(6000,10000)
%                 1       2
%
%                 3       4
\global \newcount \XqH 
\global \newcount \YqH
\global \newcount \XqB 
\global \newcount \YqB
\global \newcount \XrhoB 
\global \newcount \XrhoH
\global \XqH = 10000
\global \YqH = 9000
\drawline\photon[\SE\REG](\XqH,\YqH)[4]
%\put(\pbackx,\pbacky){\circle*{500}}  %1
\global \Xone = \pbackx
\global \Yone = \pbacky
\global \advance \pbackx by 8000
\global \Xtwo = \pbackx
\global \Ytwo = \Yone
\put(\pbackx,\pbacky){\circle*{1000}}  %2
%\drawline\gluon[\S\REG](\Xone,\Yone)[5]  %line 1 3
\global \advance \pbacky by -8000
%\put(\Xone,\pbacky){\circle*{500}}  %3
\global \Xthree = \Xone
\global \Xfour = \Xtwo
\global \newcount \YtwoU 
\global \newcount \YtwoD 
\global \YtwoU = \Ytwo 
\global \YtwoD = \Ytwo 
\global \advance  \YtwoU by 40 
\global \advance  \YtwoD by -40  
\global \Ythree = \pbacky
\global \Yfour = \Ythree
\global \newcount \YfourU 
\global \newcount \YfourD 
\global \YfourU = \Yfour 
\global \YfourD = \Yfour 
\global \advance  \YfourU by 40 
\global \advance  \YfourD by -40 
%\drawline\gluon[\S](\Xtwo,\Ytwo)[5] % line 2 4
\put(\pbackx,\pbacky){\circle*{1000}}  %4
\drawline\fermion[\E\REG](\Xtwo,\YtwoU)[4000]
\drawline\fermion[\E\REG](\Xtwo,\YtwoD)[4000]
\global \XrhoH = \pbackx
\drawline\photon[\SW\FLIPPED](\Xthree,\Ythree)[4]
\global \XqB = \pbackx
\global \YqB = \pbacky
\drawline\fermion[\E\REG](\Xfour,\YfourU)[4000]
\drawline\fermion[\E\REG](\Xfour,\YfourD)[4000]
\global \XrhoB = \pbackx
\global \newcount \XmidH 
\global \newcount \XmidB 
\global \XmidH = \Xone
\global \advance \XmidH by 4000
\global \newcount \XA
\global \XA = \Xone  %X position of quark and antiquark arrows
\global \advance \XA by 1500
\global \Xseven = \Xone
\global \advance \Xseven by 2500
\global \Xeight = \Xseven
\global \advance \Xeight by 3000
\put(\Xseven,\Yone){\oval(5000,2000)[l]}
\put(\Xseven,\Ythree){\oval(5000,2000)[l]}
\put(\Xeight,\Yone){\oval(5000,2000)[r]}
\put(\Xeight,\Ythree){\oval(5000,2000)[r]}
\global \newcount \YmidHu
\global \newcount \YmidHd
\global \newcount \YmidBu
\global \newcount \YmidBd
\global \YmidHu = \Yone
\global \YmidHd = \Yone
\global \advance \YmidHu by 1000
\global \advance \YmidHd by -1000
\global \YmidBu = \Ythree
\global \YmidBd = \Ythree
\global \advance \YmidBu by 1000
\global \advance \YmidBd by -1000
\drawarrow[\E\ATTIP](\XA,\YmidHu)
\drawarrow[\W\ATBASE](\XA,\YmidHd)
\drawarrow[\E\ATTIP](\XA,\YmidBu)
\drawarrow[\W\ATBASE](\XA,\YmidBd)
\global \advance \XA by 5500
\drawarrow[\E\ATTIP](\XA,\YmidHu)
\drawarrow[\W\ATBASE](\XA,\YmidHd)
\global \advance \YmidHu by 500
\global \advance \YmidHd by -1200
\put(\XA,\YmidHu){$\!\!\!l_1$}
\put(\XA,\YmidHd){$\!\!\!l'_1$}
\drawarrow[\E\ATTIP](\XA,\YmidBu)
\drawarrow[\W\ATBASE](\XA,\YmidBd)
\global \advance \YmidBu by 500
\global \advance \YmidBd by -1200
\put(\XA,\YmidBu){$\!\!\! l_2$}
\put(\XA,\YmidBd){$\!\!\! l'_2$}
\global \newcount \LYqH
\global \newcount \LYqB
\global \LYqH = \YqH
\global \LYqB = \YqB
\global \advance \LYqH by -1000
\global \advance \LYqB by 500
\global \newcount \LXq
\global \LXq = \XqH
\global \advance \LXq by -4000
\put(\LXq,\LYqH){$\gamma^*_L (q_1)$}
\put(\LXq,\LYqB){$\gamma^*_L (q_2)$}
\global \advance \XrhoH by 1000
\global \advance \XrhoB by 1000
\put(\XrhoH,\Yone){$\rho^0_L (k_1)$}
\put(\XrhoB,\Ythree){$\rho^0_L (k_2)$}
\global \newcount \Ycenter
\global \Ycenter = \Yone
\global \advance \Ycenter by \Ythree
\global \divide \Ycenter by 2
\put(\XmidH,\Ycenter){\oval(3000,12500)}
\global \advance \YmidHd by -3000
\drawline\fermion[\N\REG](\XmidH,\YmidHd)[2000]
\global \advance \Ycenter by 1000
\drawarrow[\N\ATTIP](\XmidH,\Ycenter)
\global \advance \Ycenter by -1000
\global \advance \XmidH by -1000
\put(\XmidH,\Ycenter){$r$}
\end{picture}
\vskip 1.5 cm
\caption{Amplitude for the process $\gamma^*_L\gamma^*_L \to \rho^0_L\rho^0_L$
in the impact representation. The blobs denote the vector meson
distribution amplitudes.
}
\label{Figprocess}
\end{figure}
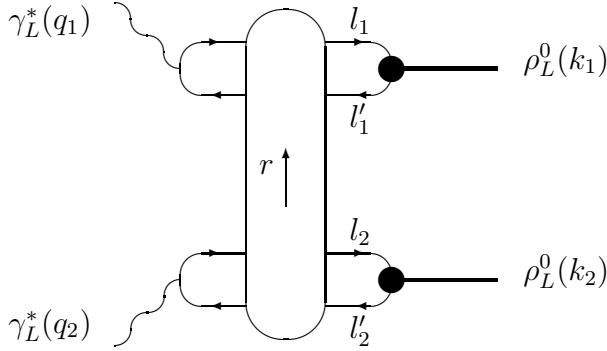
The process (\ref{process}) at high energies can be visualized
as in Fig.\ref{Figprocess}. The physical picture of this impact representation
is due to the presence of different time scales:
a $q \bar{q}$ dipole is formed from the virtual photon, then interacts by
exchanging $t-$channel gluons before recombining in a meson.
The 
Fig.\ref{Figprocess}
also explains the kinematics of the process (\ref{process}).
We parametrize the incoming photon momenta by introducing two light-like
Sudakov vectors $q'_1$ and $q'_2$ related to the incoming particles, which
satisfy 
$2 q'_1 \cdot q'_2 = s \equiv 2 q_1 \cdot q_2.$ The usual
$s_{\gamma^*\gamma^*}$ is related to the auxiliary useful variable $s$ by $s_{\gamma^*\gamma^*}=s
-Q_1^2-Q_2^2.$
In this
basis, the 
incoming photon momenta read 
\bea
\label{inmom}
&&q_1= q'_1 - \frac{Q_1^2}{s} q_2' \;,\nonumber \\  
&&q_2 =q'_2 - \frac{Q_2^2}{s} q_1'\,.
\eea~
Their polarization vectors are
\bea
\label{inpol}
&&\epsilon^{L(1)}_{\mu}= \frac{q_{1 \mu}}{Q_1} + \frac{2 Q_1}{s} q'_{2\mu}\;, 
\nonumber \\
&&\epsilon^{L(2)}_{\mu}= \frac{q_{2 \mu}}{Q_2} + \frac{2 Q_2}{s} q'_{1\mu}\;,
\eea
which is obtained after imposing the conditions $\epsilon_{L(i)}^2=1$ and
$q_i \cdot \epsilon_{L(i)}=0.$
Because of e.m. gauge invariance, polarization vectors (\ref{inpol})
can be effectively replaced by the second terms on the rhs of these
equations.
The momentum transfer in the $t-$channel is $r= k_1 -q_1.$

We label the momentum of the quarks and antiquarks entering the meson
wave functions as $l_1$ and $l'_1$ for the upper part of the diagram
and $l_2$ and $l'_2$ for the lower part (see fig.\ref{Figprocess}).

In the basis (\ref{inmom}), the vector meson momenta can be expanded in
the form
\bea
\label{rotk1k2}
&& k_1 = \alpha(k_1) \, q_1' + \frac{\rb^2}{\alpha \, s} q_2' + r_\perp \;,
\nonumber  \\
&& k_2 = \beta(k_1) \, q_2' + \frac{\rb^2}{\beta \, s} q_1' - r_\perp \,.
\eea
Note that our convention is such that for any tranverse vector
$v_\perp$ in Minkowski space, $\underline{v}$ denotes its euclidean
form. In the following, we will treat the $\rho$ meson as being massless. $\alpha$ and $\beta$
are very close to  unity, and read
\bea
\label{defalphabeta}
&&\alpha(k_1) =  \frac{1}{2} \left(1 - \frac{Q_2^2}{s}\right) \left(1 +
\sqrt{1-4 \frac{\rb^2}{s} \frac{1}{\left(1 -
\frac{Q_1^2}{s}\right)\left(1 - \frac{Q_2^2}{s}\right)}}\right)\;,
\nonumber \\
&& \beta(k_2) =  \frac{1}{2} \left(1 - \frac{Q_1^2}{s}\right) \left(1 +
\sqrt{1-4 \frac{\rb^2}{s} \frac{1}{\left(1 -
\frac{Q_1^2}{s}\right)\left(1 - \frac{Q_2^2}{s}\right)}}\right)\,.
\eea

In this decomposition, it is straightforward to relate $t=r^2$ with
$\rb^2 = -r_\perp^2$.
The corresponding relation is 
\beq
\label{tfonctionr}
t = -\frac{Q_1^2 \, Q_2^2}{s} -\frac{2 \rb^2}{1+\sqrt{1 - \frac{4
\rb^2}{s\left(1 - \frac{Q_1^2}{s}\right)\left(1 -
\frac{Q_2^2}{s}\right)}}}\left(\frac{1}{1-\frac{Q_1^2}{s}}+\frac{1}{1-\frac{Q_2^2}{s}}-1 \right) \;,
\eq
or equivalently,
\beq
\label{rfonctiont}
\rb^2 = -\left(t + \frac{Q_1^2 Q_2^2}{s}  \right)\frac{\left(1 - \frac{Q_1^2}{s}\right)\left(1 -
\frac{Q_2^2}{s}\right)}{\left(1 - \frac{Q_1^2 Q_2^2}{s^2}\right)}\left(1 +
\frac{t +\frac{Q_1^2 Q_2^2}{s}}{s \left(1 - \frac{Q_1^2 Q_2^2}{s^2}\right)}\right)\,.
\eq
From Eq.(\ref{tfonctionr}) the threshold for $|t|$ is given by
$|t|_{min}=Q_1^2 Q_2^2/s\,,$ corresponding to $r_\perp = 0.$
In the kinematical range we are interested in, the relation
(\ref{rfonctiont})
can be approximated as  $\rb^2=-t,$ in accordance with the usual result
that $r$ can be considered as purely transverse
  in the Regge limit.

We write the 
Sudakov decomposition of the quarks entering the $\rho$ mesons as
\bea
\label{l}
&& l_1 = z_1 q'_1 + l_{\perp 1} + z_1 r_\perp - \frac{(l_{\perp 1} + z_1
r_\perp)^2}{z_1 
s} q'_2 \;,\nonumber \\
&& l'_1 = \zb_1 q'_1 - l_{\perp 1} + \zb_1 r_\perp - \frac{(-l_{\perp 1} +
\zb_1
r_\perp)^2}{\zb_1 s} q'_2 \;, \nonumber \\  
&& l_2 = z_2 q'_2 + l_{\perp 2} - z_2 r_\perp - \frac{(l_{\perp 2} - z_2
r_\perp)^2}{z_2
s} q'_1 \;, \nonumber \\
&& l'_2 = \zb_2 q'_2 - l_{\perp 2} - \zb_2 r_\perp - \frac{(-l_{\perp 2} -
\zb_2
r_\perp)^2}{\zb_2 s} q'_1 \,,
\eea
where we have explicitly separated the transverse momenta $l_{\perp 1}$
($l_{\perp 2}$)  of quark and anti-quark forming the $\rho_L-$mesons 
with
respect to the
momentum $k_1$ ($k_2$).  
In the following we shall apply the collinear approximation which consists
in putting the relative momenta $l_{i\perp}$ in Eq.(\ref{l}) to zero
at  each $q \bar{q} \rho-$meson vertex. The decomposition (\ref{l}) is more
easily understood in a slightly different Sudakov basis, which up to term
linear in $r_\perp$ is obtained by the substitutions $q'_1 \to q'_1 +
r_\perp$ and $q'_2 \to q'_2 - r_\perp,$ see Eq.(\ref{rotk1k2}) with $\alpha(k_1)$
and $\beta(k_2)$ close to unity. In this new basis the $\rho$ mesons
have no transverse momenta and quarks transverse momenta have only relative momenta.

\section{Impact representation} 
\label{impact}
\setcounter{equation}{0}

\begin{figure}[htp]
\begin{picture}(4000,10000)
\global \newcount \XqH 
\global \newcount \YqH
\global \newcount \XqB 
\global \newcount \YqB
\global \newcount \XrhoB 
\global \newcount \XrhoH
\global \XqH = 10000
\global \YqH = 9000
\drawline\photon[\SE\REG](\XqH,\YqH)[4]
%\put(\pbackx,\pbacky){\circle*{500}}  %1
\global \Xone = \pbackx
\global \Yone = \pbacky
\global \advance \pbackx by 5000
\global \Xtwo = \pbackx
\global \Xfour = \Xtwo
\global \Ytwo = \Yone
\global \advance \Xone by 1000
\put(\pbackx,\pbacky){\circle*{500}}  %2
\drawline\gluon[\S\FLIPPED](\Xone,\Yone)[5]  %left gluon
%\put(\pbackx,\pbacky){\circle*{500}}  %3
\global \advance \Xone by -1000
\global \Xthree = \Xone
\global \newcount \YtwoU 
\global \newcount \YtwoD 
\global \YtwoU = \Ytwo 
\global \YtwoD = \Ytwo 
\global \advance  \YtwoU by 40 
\global \advance  \YtwoD by -40  
\global \Ythree = \pbacky
\global \Yfour = \Ythree
\global \newcount \YfourU 
\global \newcount \YfourD 
\global \YfourU = \Yfour 
\global \YfourD = \Yfour 
\global \advance  \YfourU by 40 
\global \advance  \YfourD by -40 
\global \advance \Xone by 4000
\drawline\gluon[\S\REG](\Xone,\Ytwo)[5] %right gluon
\global \advance \Xone by -4000
\put(\Xfour,\pbacky){\circle*{500}}  %4
\drawline\fermion[\E\REG](\Xtwo,\YtwoU)[4000]
\drawline\fermion[\E\REG](\Xtwo,\YtwoD)[4000]
\global \XrhoH = \pbackx
\drawline\photon[\SW\FLIPPED](\Xthree,\Ythree)[4]
\global \XqB = \pbackx
\global \YqB = \pbacky
\drawline\fermion[\E\REG](\Xfour,\YfourU)[4000]
\drawline\fermion[\E\REG](\Xfour,\YfourD)[4000]
\global \XrhoB = \pbackx
\global \newcount \XmidH 
\global \newcount \XmidB 
\global \XmidH = \Xone
\global \advance \XmidH by \Xtwo
\global \divide \XmidH by 2
\global \XmidB = \Xthree
\global \advance \XmidB by \Xfour
\global \divide \XmidB by 2
\put(\XmidH,\Yone){\oval(5000,2000)}
\put(\XmidH,\Ythree){\oval(5000,2000)}
\global \newcount \YmidHu
\global \newcount \YmidHd
\global \newcount \YmidBu
\global \newcount \YmidBd
\global \YmidHu = \Yone
\global \YmidHd = \Yone
\global \advance \YmidHu by 1000
\global \advance \YmidHd by -1000
\global \YmidBu = \Ythree
\global \YmidBd = \Ythree
\global \advance \YmidBu by 1000
\global \advance \YmidBd by -1000
\drawarrow[\E\ATTIP](\XmidH,\YmidHu)
\drawarrow[\W\ATBASE](\XmidH,\YmidHd)
\drawarrow[\E\ATTIP](\XmidB,\YmidBu)
\drawarrow[\W\ATBASE](\XmidB,\YmidBd)
\global \newcount \LYqH
\global \newcount \LYqB
\global \LYqH = \YqH
\global \LYqB = \YqB
\global \advance \LYqH by -1000
\global \advance \LYqB by 500
\global \newcount \LXqH
\global \newcount \LXqB
\global \LXqH = \XqH
\global \LXqB = \XqB
\global \advance \LXqH by -4000
\global \advance \LXqB by -4000
\put(\LXqH,\LYqH){$\gamma^*_L (q_1)$}
\put(\LXqB,\LYqB){$\gamma^*_L (q_2)$}
\global \advance \XrhoH by 1000
\global \advance \XrhoB by 1000
\put(\XrhoH,\Yone){$\rho^0_L (k_1)$}
\put(\XrhoB,\Ythree){$\rho^0_L (k_2)$}
\global \advance \Yone by -2000
\global \advance \Xone by 1000
\drawarrow[\N\ATTIP](\Xone,\Yone)
\global \advance \Xone by 3000
\drawarrow[\N\ATTIP](\Xone,\Yone)
\global \advance \Xone by -4000
\global \advance \Yone by -1000
%\global \advance \Xone by 4000
\put(\Xone,\Yone){$\! k$}
\put(\Xtwo,\Yone){$\! \!  r-k$}
\end{picture}
\vskip 0.5 cm
\caption{Amplitude for the process $\gamma^*_L\gamma^*_L \to \rho^0_L\rho^0_L$
at Born order. The $t$-channel gluons are attached to the quark lines in all
possible ways.}
\label{Figborn}
\end{figure}
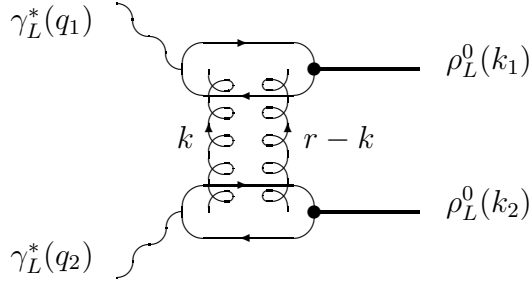

The impact representation of the scattering amplitude for the process
(\ref{process}) has the form (see Fig.\ref{Figborn})

% Fig.2    
%                 1       2
%
%                 3       4

\be 
\label{M}
{\cal M} = is\;\int\;\frac{d^2\,\kb}{(2\pi)^4\kb^2\,(\rb -\kb)^2}
{\cal J}^{\gamma^*_L(q_1) \to \rho^0_L(k_1)}(\kb,\rb -\kb)\;
{\cal J}^{\gamma^*_L(q_2) \to \rho^0_L(k_2)}(-\kb,-\rb +\kb)\,,
\ee
where ${\cal J}^{\gamma^*_L(q_1) \to \rho^0_L(k_1)}(\kb,\rb -\kb)$   
(${\cal J}^{\gamma^*_L(q_2) \to \rho^0_L(k_2)}(\kb,\rb -\kb)$)
are the impact factors corresponding  to the                            
transition of
$\gamma^*_L(q_1)\to \rho^0_L(k_1)$ ($\gamma^*_L(q_2)\to \rho^0_L(k_2)$) via the
$t-$channel exchange of two gluons. The impact factors are the
$s-$channel discontinuities of the corresponding $S-$matrices describing
the $\gamma^*_L \;g\to
\rho^0_L\; g$ processes projected on the longitudinal (nonsense) polarizations
of the virtual gluons in the $t-$channel
\beq
\label{if}
{\cal J}^{\gamma^*_L(q_1) \to \rho^0_L(k_1)}(\kb,\rb -\kb)= \int
\frac{d\beta}{s}
S^{\gamma^*_L(q_1) \to \rho^0_L(k_1)}_{\mu\nu}q_2^{'\mu} q_2^{'\nu}\;,\nonumber
\ee
\beq
%\label{if}
{\cal J}^{\gamma^*_L(q_2)    \to \rho^0_L(k_2)}(-\kb,-\rb +\kb)= \int
\frac{d\alpha}{s}
S^{\gamma^*_L(q_2) \to \rho^0_L(k_2)}_{\mu\nu}q_1^{'\mu} q_1^{'\nu}\,.\nonumber
\ee
Here the integration variables $\alpha $ and $\beta$ are defined by the
Sudakov decomposition of the gluonic momentum $k$
\be
\label{sudk}
k= \alpha q_1' + \beta q_2' + k_\perp\;.
\ee
The amplitude (\ref{M}) depends linearly on $s$, since these impact factors
are $s$-independent.  
Calculations of the impact factors in the Born approximation are standard
\cite{ginzburg}. \footnote{Recently the forward impact factor of $\gamma^*_L(Q^2) \to \rho^0_L$ 
transition was calculated at the next-to-leading order accuracy in Ref.
\cite{ivanov}.}
 They read
\bea
\label{ifup}
&&\hspace{-.8cm}{\cal J}^{\gamma^*_L(q_1) \to \rho_L(k_1)}(\kb,\rb -\kb)
\nonumber \\
&&\hspace{-.8cm}= \int\limits_0^1 dz_1 z_1 \, \zb_1 \, \phi(z_1)8 \pi^2 \alpha_s
\frac{e}{\sqrt{2}} \frac{\delta^{a b}}{2 N_c} Q_1 \, f_\rho \alpha(k_1)
\rm{P_P(z_1,\kb,\rb,\mu_1)} 
\eea
and
\bea
\label{ifdown}
&&\hspace{-.8cm}{\cal J}^{\gamma^*_L(q_2) \to \rho^0_L(k_2)}(-\kb,-\rb +\kb)
\nonumber \\
&&\hspace{-.8cm}= \int\limits_0^1 dz_2 z_2 \, \zb_2 \, \phi(z_2)8 \pi^2 \alpha_s
\frac{e}{\sqrt{2}} \frac{\delta^{a b}}{2 N_c} Q_2 \, f_\rho \beta(k_2)
\rm{P_P(z_2,\kb,\rb,\mu_2)}\;,
\eea
where 
\beq
\label{defPP}
\rm{P_P(z_1,\kb,\rb,\mu_1)}=
 \frac{1}{z_1^2\rb^2 + \mu_1^2} +
\frac{1}{\zb_1^2\rb^2 + \mu_1^2} 
 - \frac{1}{(z_1\rb -\kb)^2 + \mu_1^2} - \frac{1}{(\zb_1\rb
-\kb)^2
+ \mu_1^2} 
\eq
is proportional to the impact factor of quark pair production from a
longitudinally polarized photon, with two t-channel exchanged gluons.
In formula (\ref{defPP}) the collinear approximation $l_\perp=0$ has 
been made.

In this expression $\mu_1^2=Q_1^2 \; z_1 \; \zb_1 + m^2$ and 
$\mu_2^2=Q_2^2 \; z_2 \; \zb_2 + m^2,$ where $m$ is the quark mass.
The limit $m \to 0$ is regular and we will restrict ourselves to
the light quark case, taking thus $m=0.$ 
Note that Eq.(\ref{ifdown}) can be obtained from Eq.(\ref{ifup}) from
 the combination of two substitutions, firstly $(z_1, \, Q_1) \to (z_2,
 \, Q_2)$ 
and secondly $(\kb,\, \rb)
\to (\kb -\rb,\, -\rb).$ This second substitution is easy to understand
since for the upper (lower) blob, the total $t-$channel outgoing momentum is
$r_\perp$ $(-r_\perp),$ the outgoing gluon carries momentum $k$ ($k-r$),
and the incoming gluon carries momentum $k-r$ ($k$).
This substitution effectively corresponds to exchanging the third and
the fourth term in Eq. (\ref{defPP}), and leaves this impact factor
 invariant,
due to its symmetry under $z \to \zb,$ which is reminiscent of the Pomeron
structure of the t-channel state (namely, the impact factor is even
 under C conjugation).
Thus, only the first substitution is necessary to obtain the upper blob
from the lower blob.

In the formulae (\ref{ifup}, \ref{ifdown}), $\phi$ is the distribution
amplitude of the produced longitudinally polarized
$\rho-$mesons. For the case with quark $q$ of one flavour
it is defined (see, e.g. \cite{BB}) 
by the matrix element of non-local, gauge invariant correlator
of quark fields on the light-cone
\begin{equation}
\label{da}
\langle 0| \bar q(x)\;\gamma^\mu\;q(-x)|\rho_L(p)\rangle = f_\rho \;p^\mu 
\int\limits_0^1 dz e^{i(2z-1)(px)}\phi(z)\;,
\end{equation}
where the coupling constant is $f_\rho=216$ MeV and where the gauge links 
are omitted to simplify the notation. The amplitudes for production of 
$\rho^0$'s are obtained by noting that $|\rho^0\rangle = 
1/\sqrt{2}(|\bar u u\rangle -|\bar d d \rangle )$.

The structures in the formulae (\ref{ifup}, \ref{ifdown}) arise
 due to the collinear approximation. Indeed, one can 
neglect the $l_\perp$ dependencies in the propagators, when computing the impact factors.
Thus the only dependency with respect to this transverse momentum of the quark, in the Sudakov basis of the $\rho-$meson, is inside the wave function of these $\rho-$mesons. 
When integrating over the phase space of the quarks and antiquarks, this results
in integrating the meson wave functions up to the factorization scale (which is of the order
of the photon virtualities). We neglect here any evolution with respect
to this scale.
Such integrated wave function 
 is by definition the distribution amplitude
of the $\rho-$mesons. For simplicity, we use the asymptotic distribution amplitude
\beq
\phi(z)=6 \, z \, (1-z)\,.
\eq
The only remaining part the quark phase space integrations are the integration with respect
to quark longitudinal fractions of meson momenta $z_1$ and $z_2.$

Combining Eqs.(\ref{M}, \ref{ifup}, \ref{ifdown}, \ref{defPP}), the
amplitude can be expressed
as
\beq
\label{MCalgeneral}
{\cal M} = i\, s\, 2 \,\pi \,\frac{N_c^2-1}{N_c^2}\, \alpha_s^2 \,\alpha_{em}
\alpha(k_1)\, \beta(k_2)\, f_\rho^2\, Q_1\, Q_2 \, \intfeyn d z_1 \, d z_2 \, z_1
\, \zb_1 \, \phi(z_1)\, z_2
\, \zb_2 \, \phi(z_2) {\rm M}(z_1,\, z_2)\,,
\eq
with 
\bea
\label{defM}
&& \hspace{-.8cm}{\rm M}(z_1,\, z_2)=\int \frac{d^2 \kb}{\kb^2 (\rb-\kb)^2} \left[ \frac{1}{z_1^2\rb^2 + \mu_1^2} +
\frac{1}{\zb_1^2\rb^2 + \mu_1^2} 
 - \frac{1}{(z_1\rb -\kb)^2 + \mu_1^2} - \frac{1}{(\zb_1\rb
-\kb)^2
+ \mu_1^2} \right] \nonumber \\
&& \hspace{2cm}\times \left[ \frac{1}{z_2^2\rb^2 + \mu_2^2} +
\frac{1}{\zb_2^2\rb^2 + \mu_2^2}
 - \frac{1}{(z_2\rb -\kb)^2 + \mu_2^2} - \frac{1}{(\zb_2\rb
-\kb)^2
+ \mu_2^2} \right]\,.
\eea
In term of this amplitude, the differential cross-section can be expressed
in the large $s$ limit (neglecting terms of order $Q_i^2/s$) as
\beq
\label{crosssection}
\frac{d \sigma^{\gamma^*_L \gamma^*_L \to \rho^0_L \rho^0_L}}{dt}=\frac{|{\cal M}|^2}{16
\, \pi \,s^2 }\,.
\eq

\section {Cross section at Born order}
\label{cross}
\setcounter{equation}{0}

\subsection{Forward case}

We begin with the simpler case $t= t_{min}$ ({\it i.e.}
 $\rb=0$), where the final result for the function $M(z_1,z_2)$ (\ref{defM}) can be written
in a rather simple form.
The integral over $\kb$ can readily be performed and gives
\beq
\label{Mlimite}
M(z_1,z_2) = \frac{4 \pi}{z_1 \, \zb_1 z_2 \, \zb_2 Q_1^2 \, Q_2^2\,
(z_1 \, \zb_1 Q_1^2 - z_2 \, \zb_2 Q_2^2)} \ln \frac{z_1 \, \zb_1 Q_1^2}{z_2 \, \zb_2 Q_2^2}\,.
\eq
The amplitude ${\cal M}$ given by Eq.(\ref{MCalgeneral}) can then be computed
analytically from Eq.(\ref{Mlimite}) through double integration over  $z_1$ and $z_2$. 
This is explained in appendix \ref{tmin}.
It results in
\bea
\label{resultMmin}
&&\hspace{-1.1cm}{\cal M}_{t_{min}}= -i s \, \frac{N_c^2-1}{N_c^2} \, \alpha_s^2 \, \alpha_{em} \,
\alpha(k_1) \, \beta(k_2) \, f_\rho^2 \, \frac{9 \pi^2}{2} \, \frac{1}{Q_1^2
  Q_2^2}  \left[6 \, \left(R + \frac{1}{R}\right) \ln^2 R  \right.\\
&&\hspace{-1cm}\left. + \,12\,
\left(R-\frac{1}{R}\right) \ln R \,+\, 12 \,  \left(R + \frac{1}{R}\right)\,+\,
\left(3 \,R^2 +\, 2\,+\frac{3}{R^2}\right) \left(\,\ln\,(1-R)\,\ln^2 R
  \nonumber \right. \right.\\
&&\hspace{-1cm}\left. \left.-\ln \,(R+1)\,\ln^2 R \,-\,2 \,{\rm Li_2}\,(-R)\,
    \ln R \,+\, 2\,{\rm Li_2}\,(R)\, \ln R \,+\,2\,{\rm Li_3}\,(-R)\,-\,2\, {\rm Li_3}\,(R)\right)\right]\,,\nonumber
\eea
where $R=Q_1/Q_2.$

In the special case where $Q=Q_1=Q_2,$ it simplifies immediately to 
\beq
\label{amplitudeR1}
{\cal M}_{t_{\min}}(Q_1=Q_2) \sim \, -i \,s \,  \frac{N_c^2-1}{N_c^2} \, \alpha_s^2 \, \alpha_{em} \,
\alpha(k_1) \, \beta(k_2) \, f_\rho^2 \, \frac{9 \pi^2}{2 Q^4}\, (24 - 28 \, \zeta(3))\,.
\eq
The peculiar limits $R \gg 1$ and $R \ll 1$ are of special physical interest,
since they correspond to the kinematics typical for deep inelastic scattering
on a photon target described through collinear approximation, {\it i.e.} the usual
parton model. At moderate values of $s,$ apart from diagrams with gluon
exchange
in $t-$channel considered here, one should also take into account of diagrams 
with quark exchange. We do not consider them here since we restrict ourselves
to the asymptotical region of large $s.$ 

In the limit $R \gg 1$, the amplitude simplifies into
\beq
\label{amplitudeparton}
{\cal M}_{t_{\min}} \sim i s \frac{N_c^2-1}{N_c^2} \, \alpha_s^2 \, \alpha_{em} \,
\alpha(k_1) \, \beta(k_2) \, f_\rho^2 \, \frac{96 \pi^2}{Q_1^2\, Q_2^2}
\left(\frac{\ln R}{R}-\frac{1}{6 R} \right)\,.
\eq
This result can be obtained directly by imposing from the very beginning 
the $k_\perp$ ordering typical of parton model. This is shown explicitly in
Appendix \ref{tminparton}.

\subsection{Non forward case}

In this section we will compute the amplitude (\ref{M}),  where we have been
able to perform analytically  the $k_\perp$
integrals. It involves the evaluation of a box diagram with two distinct massive
propagators and two massless propagators (denoted $I_{4 m_a m_b}$ below). We are not aware of any previous
analytic calculation of such an integral. This gives us the possibility of
studying various kinematical limits in variables $Q_1 ^2, \, Q_2 ^2, \,t$.

$M(z_1,\,z_2)$ as defined in Eq.(\ref{defM}) is symmetric under $z_1
\lr \zb_1$ and under $z_2
\lr \zb_2.$ Since on the other hand
 the distribution amplitude $\phi(z)$ is also symmetric under $z
\lr \zb,$ $M(z_1, \, z_2)$ can be modified by adding any
antisymmetric
term, 
since the integration over $z_1$ and $z_2$ will
automatically select its symmetric part.

One thus writes
\beq
\label{Masymmetric}
{\cal M}= i s 2 \pi \frac{N_c^2-1}{N_c} \alpha_s^2 \alpha_{em}
\alpha(k_1) \beta(k_2) f_\rho^2 Q_1 Q_2  \intfeyn d z_1 \, d z_2 \, z_1
\, \zb_1 \, \phi(z_1)\, z_2
\, \zb_2 \, \phi(z_2) \,  M_{asym.}(z_1,\, z_2)\;,
\eq
with
\bea
\label{Mcompact}
&&\hspace{-.7cm} M_{asym.}(z_1,\, z_2) = \int \frac{d^2 \kb}{\kb^2 (\rb-\kb)^2} 
\left[ \frac{1}{(z_1^2 \rb^2 + \mu_1^2)(z_2^2 \rb^2 + \mu_2^2)}  -\frac{1}{(z_1^2 \rb^2 + \mu_1^2)((z_2 \rb-\kb)^2 + \mu_2^2)} \right. \nonumber \\
&& \left. -\frac{1}{(z_2^2 \rb^2 + \mu_2^2)((z_1 \rb-\kb)^2 + \mu_1^2)}
+\frac{1}{((z_1 \rb-\kb)^2 + \mu_1^2)((z_2 \rb-\kb)^2 + \mu_2^2)} \right]\,.
\eea
$M_{asym.}(z_1,\, z_2)$ can be expressed in term 
of three kind of integrals, namely,

\beq
\label{I2}
I_2=\int \frac{d^d \kb}{\kb^2 (\kb - \pb)^2}\,,
\eq

\beq
\label{I3m}
I_{3m}=\int \frac{d^d \kb}{\kb^2 (\kb - \pb)^2((\kb-\ab)^2 + m^2)}\,,
\eq

and
\beq
\label{I4mm}
I_{4m_am_b}=\int \frac{d^d \kb}{\kb^2 (\kb - \pb)^2((\kb-\ab)^2 + m_a^2)((\kb-\bb)^2 + m_b^2)}\,,
\eq
where we use the dimensional regularization $d = 2 +2 \epsilon$. 
However, since we will effectively rely, for computing the
various integrals involved,  on a method which
is applicable only for both UV and IR finite integrals, it is more
efficient
to rewrite ${\cal M}$ (\ref{MCalgeneral}) in terms of another asymmetric two dimensional amplitude of the
form
\bea
\label{Mcompact2}
&& \tilde{M}(z_1,\, z_2) = - \left( \frac{1}{z_1^2 \rb^2 +\mu_1^2}
+\frac{1}{\zb_1^2 \rb^2 +\mu_2^2}\right) J_{3 \mu_2}(z_2) -( 1 \lr 2) 
\nonumber \\
&&+J_{4 \mu_1 \mu_2}(z_1,\, z_2) + J_{4 \mu_1 \mu_2}(\zb_1,\, z_2)\;,
\eea
where 
\beq
\label{defJ3m}
J_{3 \mu}(a)= \int \frac{d^2 \kb}{\kb^2 (\kb - \rb)^2} \left
[ \frac{1}{(\kb - \rb a)^2 + \mu^2} -\frac{1}{a^2 \rb^2 + \mu^2}+ (a \lr
\ab) \right] \,,
\eq
and 
\bea
\label{defJ4mm}
&&J_{4 \mu_1 \mu_2}(z_1,\, z_2)= \int \frac{d^2 \kb}{\kb^2
(\kb - \rb)^2} \nonumber \\
&& \hspace{-1cm}\times \left[ \frac{1}{((\kb - \rb z_1)^2 + \mu_1^2)((\kb - \rb z_2)^2 + \mu_2^2)}
-\frac{1}{(z_1^2 \rb^2 + \mu_1^2)(z_2^2 \rb^2 +\mu_2^2)}+ (z \lr \zb) \right] \,.
\eea
$J_{3 \mu}$ and $J_{4 \mu_1 \mu_2}$ are two dimensional integrals with
respectively 3 propagators (1 massive) and 4 propagators (2 massive,
with different masses). They are both IR and UV finite. Their computation
by brute force technique, using Feynman parametrization, seems untractable
 in such a form (specially for $J_{4 \mu_1 \mu_2}$). However, applying
a trick inspired by conformal field theories, it is possible to compute these integrals. 
The basic idea is to perform special conformal inverse transformations,
considered here in
momemtum space.
Although these two integrals, because of mass terms, are not conformal
invariant, this is actually efficient after a suitable
redefinition of the massive parameters. This is presented
 in the appendix \ref{integrals}.
\\

To complete the evaluation of the amplitude ${\cal M},$ one needs to integrate
over the quark momentum fractions in the $\rho$ mesons $z_1$ and $z_2.$
In the general case, for arbitrary values of $t,$ it seems not possible to
perform the $z_1$ and $z_2$ integrations analytically. 
We thus do them numerically.
In course of them, we observe the absence of end-point singularity when $z_{1(2)} \to 0$ or
$z_{1(2)} \to 1,$ since $P_P$ as defined in Eq.(\ref{defPP}) diverges like
$1/z,$ $1/(1-z)$ when $z \to 0,\, 1.$ This leads to perfectly stable numerical integrations.

\subsection{Results}
\label{results}

We use now the previous formulae in order to get prediction for
production rate of diffractive double  $\rho$ production. 
Running of $\alpha_S$ is a subleading effect with respect to our treatment.
Anyway, we choose to replace $\alpha_S^2$ in the various formulae presented
above by $\alpha_S(Q_1^2) \, \alpha_S(Q_2^2)$ in order to fix the coupling,
and use the three-loop running coupling $\alpha_S(Q_1^2)$ and
$\alpha_S(Q_2^2)$ with $\Lambda_{\overline{MS}}^{(4)}= 305 \,$MeV 
(see, e.g. \cite{bethke}). 
%The value of $f_{\rho}$ is taken to be $216 \,  {\rm MeV.}$

\begin{figure}[htb]
\begin{picture}(10000,33200)
\put(-1000,2000){\epsfxsize=15.7cm{\centerline{\epsfbox{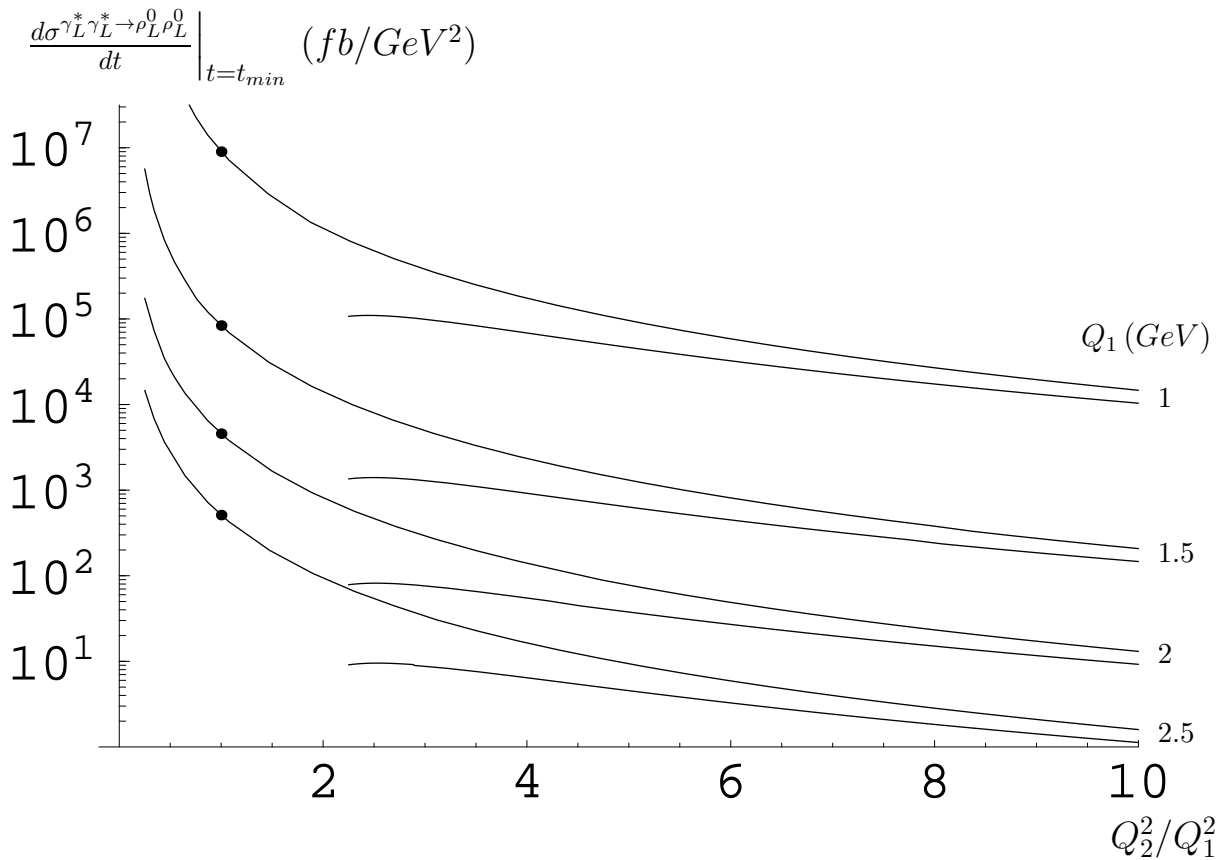}}}}
\put(0,31000){ \large $\left.\frac{d \sigma^{\gamma^*_L\gamma^*_L\to \rho^0_L
      \rho^0_L}}{d t}\right|_{t=t_{min}} \, (fb/GeV^2)$}
\put(41500,1000){\large $Q_2^2/Q_1^2$}
\put(40000,20000){ $Q_1 \, (GeV)$}
\put(43300,17700){\small $1$}
\put(43300,11800){\small $1.5$}
\put(43300,8000){\small $2$}
\put(43300,5000){\small $2.5$}
\end{picture}
\caption{Differential cross-section
 for the process $\gamma^*_L\gamma^*_L \to \rho^0_L\rho^0_L$
at Born order, at the threshold $t=t_{min},$ as a function of $Q_2^2/Q_1^2.$
The dots represents the value of the cross-section at the special point
$Q_1=Q_2,$ 
as given by the analytical formula (\ref{amplitudeR1}). The
asymptotical curves are valid for large $Q_2^2/Q_1^2,$ as predicted by
the asymptotical form (\ref{amplitudeparton})}.
\label{Figtmin}
\end{figure}

In Fig. \ref{Figtmin},
 we display the differential cross-section $d \sigma(\gamma^*_L\gamma^*_L\to \rho_L^0
      \rho_L^0)/d t$ for vanishing transverse $t-$channel
      momentum, {\it i.e.} $t=t_{min},$ as a
function of the ratio $Q_2^2/Q_1^2.$ Curves are labelled by the values of
      $Q_1.$ 
The dots on the curves represent the values of the cross-section at the special point
$Q_1=Q_2,$ which obviously correspond to the analytical formula (\ref{amplitudeR1}).
The cross-section dramatically decreases when $Q_2^2/Q_1^2$ increases  at
      fixed $Q_1.$ For comparison, we show for each value of $Q_1$ the
      asymptotical curve 
obtained by combining Eq.(\ref{resultMmin})  with
Eq.(\ref{crosssection}). The complete result approaches quickly its
      asymptotical curve. However, in view of the strong decrease of the differential
      cross-section with increasing $Q_2^2/Q_1^2,$ the asymptotical result
seems to be of little interest for estimating data rates in the most favorable kinematics.

\begin{figure}[htb]
\begin{picture}(8000,32100)
\put(41000,11000){ $Q \, (GeV)$}
\put(43500,9500){\small $1$}
\put(43500,7800){\small $1.25$}
\put(43500,6700){\small $1.5$}
\put(43500,5700){\small $1.75$}
\put(43500,4800){\small $2$}
\put(43500,3900){\small $2.25$}
\put(43500,3000){\small $2.5$}
\put(40000,-500){$-t \, (GeV^2)$}
\put(100,30000){\large $\frac{d \sigma^{\gamma^*_L\gamma^*_L\to \rho^0_L \rho^0_L}}{d t} \, (fb/GeV^2)$}
\put(-600,1000){\epsfxsize=15.7cm{\centerline{\epsfbox{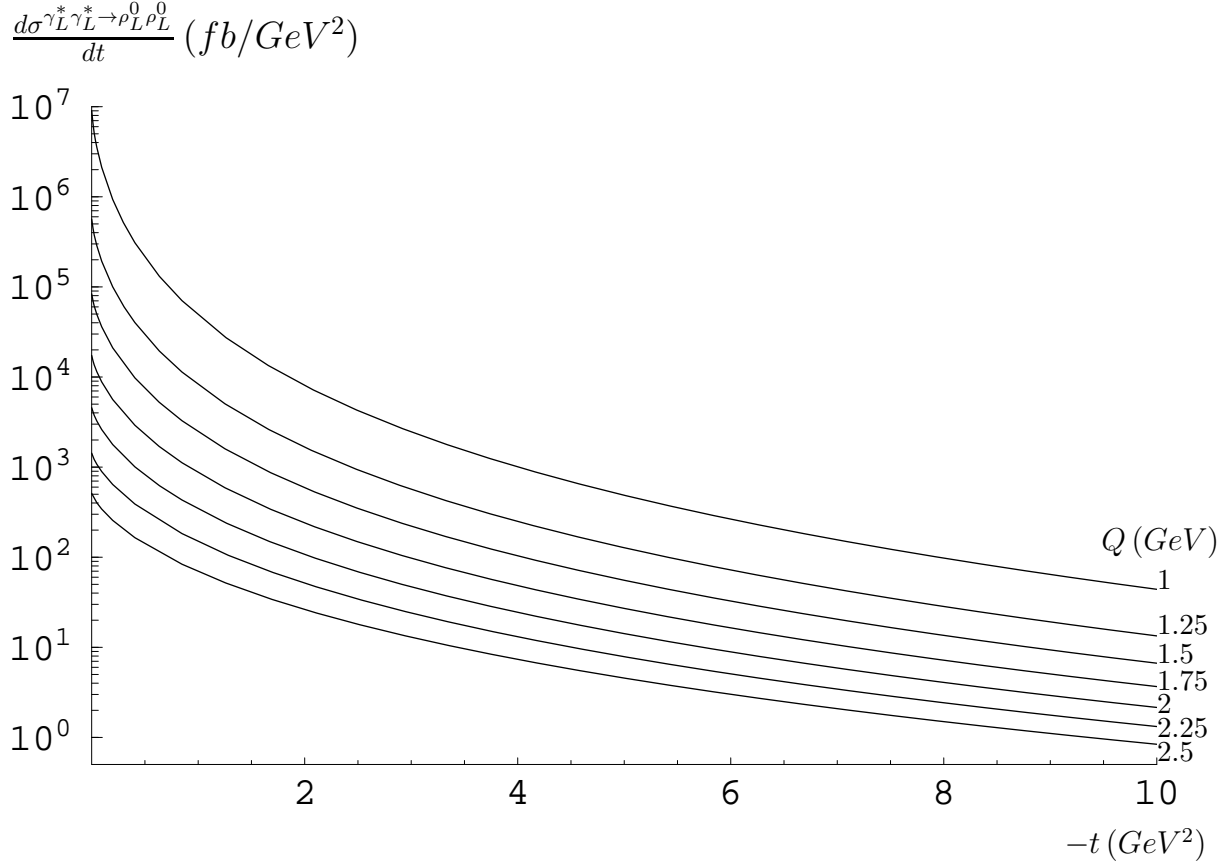}}}}
\end{picture}
\caption{Differential cross-section
 for the process $\gamma^*_L\gamma^*_L \to \rho^0_L\rho^0_L$
at Born order, as a function of $t,$ for different values of $Q = Q_1=Q_2.$}
\label{Figdsdt}
\end{figure}

The $t$-dependence of the differential cross-section $d \sigma/dt$ 
is displayed in Fig. \ref{Figdsdt} for various values of $Q=Q_1=Q_2.$

As anticipated, the cross-section is strongly peaked in the forward direction.
This fact is less dangerous than for the real photon case since the virtual
 photon is not in the direction of the beam. 
However, the differential cross-section seems to be sufficient for the
$t-$dependence to be measured up to a few ${\rm GeV^2.}$
The comparison of the curves on Fig.\ref{Figtmin} for $t=t_{min}$ with those
on Fig.\ref{Figdsdt}
leads to the conclusion that the phenomenological predictions obtained in the
forward case will practically dictate the general trends of integrated cross-sections.

\begin{figure}[htb]
\begin{picture}(8000,33700)
\put(0,2000){\epsfxsize=16.3cm{\centerline{\epsfbox{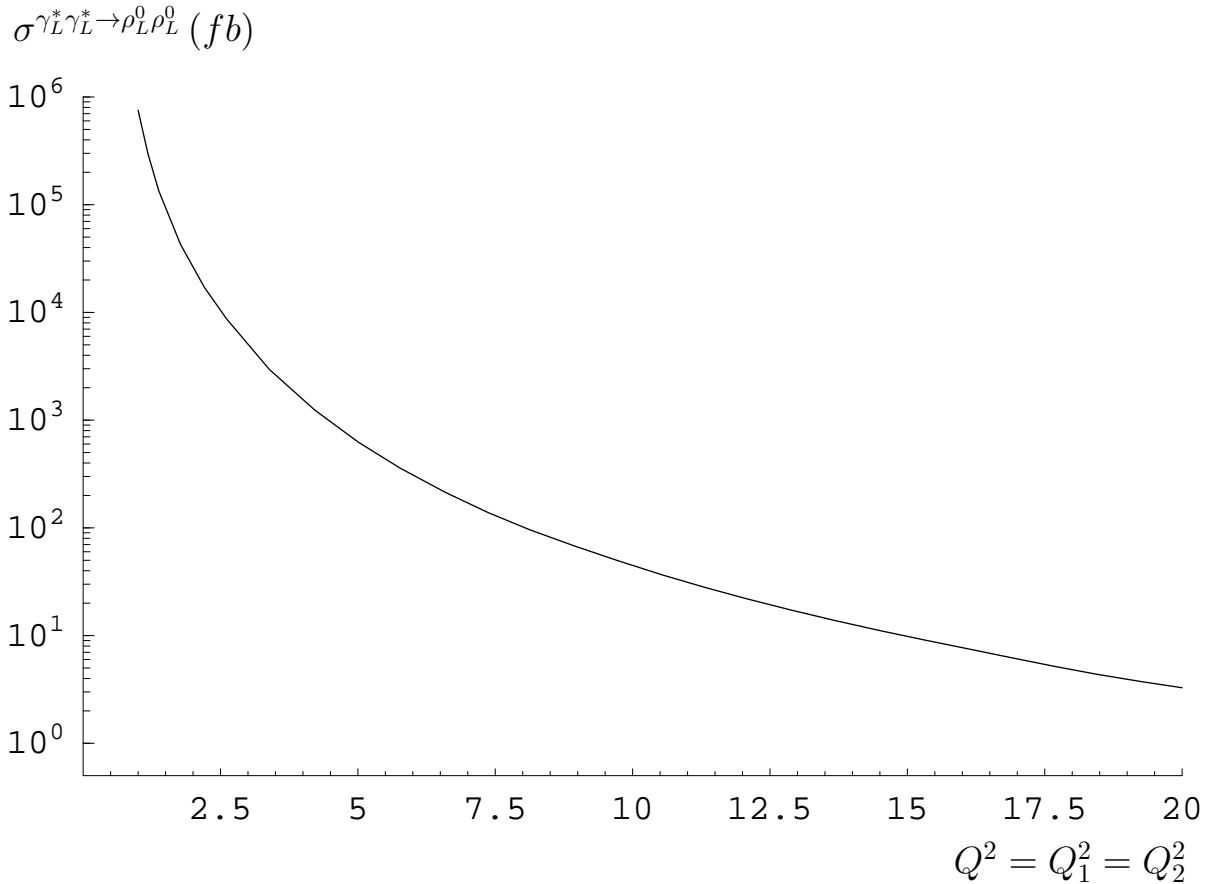}}}}
\put(0,32000){ \large $\sigma^{\gamma^*_L\gamma^*_L\to \rho^0_L \rho^0_L} \, (fb)$}
\put(36000,500){\large $Q^2=Q_1^2=Q_2^2$}
\end{picture}
\caption{The integrated cross-section
 for the process $\gamma^*_L\gamma^*_L \to \rho^0_L\rho^0_L$
at Born order as a function of $Q_1^2=Q_2^2.$}
\label{FigsigmaQ2}
\end{figure}

Figure \ref{FigsigmaQ2} shows the integrated over $t$ cross-section as a
function of $Q^2=Q_1^2=Q_2^2.$ The magnitude of the cross-section seems to be
sufficient for a detailed study to be performed at the linear collider presently
under study. Note that we did not multiply by the virtual photon fluxes, which
would amplify the dominance of smaller $Q^2.$  However, triggering efficiency
often increases substantially with $Q^2$ \cite{royon}.
At this level of calculation there is no $s$-dependence of the
cross-section. It will appear after taking into account of BFKL evolution. 

\section{Conclusion} 
This Born order study shows that the process  $\gamma^*(Q_1^2)\gamma^*(Q_2^2) \to
\rho_L^0 \rho_L^0$ can be measured  at foreseen $e^+-e^--$colliders for
$Q_1^2, Q_2^2$ up to a few ${\rm GeV^2}.$ 
 Indeed, a nominal integrated luminosity of $100 {\rm fb}^{-1}$ should
yield thousands of events per year, with ${ Q^2 \gtrsim 1}$ GeV$^2$.
 This would open 
a new domain of investigation for diffractive processes in which practically
all
ingredients of the scattering amplitude are under control within a pertubative
approach.

In the near future we expect to
include the transversely polarized photon contribution, which should slightly 
enhance 
the non forward amplitude (this amplitude obviously vanishes at $t=t_{min}$).
We also intend to incorporate the effect of BFKL evolution, since
  resummation effects are expected to give a net and visible enhancement of
  the cross section. 

Finally, let us note that the elusive Odderon may also be looked for in
$\gamma^*  \gamma^*$ exclusive reactions \cite {ewerz}, 
and that one may use the strategy developed in Ref.\cite{Hagler} to find it
through its interference with Pomeron exchange which gives rise to 
charge asymmetries in $\gamma^* \gamma^* -> \pi \pi \pi \pi.$
\\
\vskip.1in
\noindent
NOTE ADDED : After this paper has been accepted, two works \cite{new}
%confirm and 
improve our
analysis by taking into account of higher order effects.
\vskip.1in
\noindent
{\large \bf Acknowledgements:}
\\

 We thank R.~Enberg, N.~Kivel, S.~Munier, M.~Segond and A.~Shuvaev for
discussions. This  work  is  supported  by  
the Polish Grant 1 P03B 028 28, 
the
French-Polish scientific agreement Polonium.
%and the Joint Research
%Activity "Generalised Parton Distributions" of the european I3 program
%Hadronic Physics, contract RII3-CT-2004-506078. 
L.Sz. is a Visiting Fellow of 
the Fonds National pour la Recherche Scientifique (Belgium).

%%%%%%%%%%%%%%%%%%%%%%%%%%%%%%%%%%%%%%%%%%%%%%%%%%%%%%%%%%%%%%%%%%%%%%%%%%%%%%

\begin{appendix}
\section{Appendices}

\setcounter{equation}{0}

\subsection{Computation of the amplitude at $t=t_{min}$}
\label{tmin}

In this appendix we will prove the result (\ref{resultMmin}) for the
production amplitude at $|t|_{min}=Q_1^2 Q_2^2/s\,,$ still neglecting the $\rho$-mass.
The amplitude (\ref{MCalgeneral}) can be written as 
\beq
\label{MCalthreshold}
{\cal M}=\,i s \, \frac{N_c^2-1}{N_c^2} \, \alpha_s^2 \, \alpha_{em} \,
\alpha(k_1) \, \beta(k_2) \, f_\rho^2 \, \frac{288 \pi^2}{Q_1
  Q_2} \,K\,,
\eq
where
\beq
\label{defK}
K= \intfeyn d z_1 \,\intfeyn d z_2 \, \int\limits_0^\infty \, d k^2 \frac{z_1 \, \zb_1 z_2 \, \zb_2}{(k^2+z_1 \, \zb_1 Q_1^2)(k^2 + z_2 \, \zb_2 Q_2^2)} \,,
\eq 
which reduces to 
\beq
\label{resK}
K= \intfeyn d z_1 \,\intfeyn \, d z_2 \, \frac{z_1 \, \zb_1 z_2 \, \zb_2}{z_1 \, \zb_1 Q_1^2 - z_2 \, \zb_2 Q_2^2} \ln \frac{z_1 \, \zb_1 Q_1^2}{z_2 \, \zb_2 Q_2^2}\,.
\eq 
In the following, we denote $R=Q_1/Q_2.$
We first reduce the computation of $K$ to a one dimensional integral evaluation. 
Performing the change of variables $x_1=4 \, z_1 \, \zb_1$ and
 $x_2=4 \, z_2 \, \zb_2,$ $K$ reads
\beq
\label{K1}
K\,=\, \frac{1}{4^2 Q_1^2}\intfeyn  \frac{d x_1}{\sqrt{1-x_1}} \, \frac{d
  x_2 \, x_2}{\sqrt{1-x_2}}\, \left(1-  \frac{1}{1-
  R^2\,x_1/x_2}\right)\,\ln \left(R^2 \,\frac{x_1}{x_2}\right)\,.
\eq 
After replacing the variable $x_1$ by $x=x_1/x_2,$ $K$ reads
\beq
\label{K2}
K \,= \, \frac{1}{4^2 Q_1^2}\intfeyn  \frac{d x_2 \, x_2^2}{\sqrt{1-x_2}} \,
  \int\limits_0^{\, \quad 1/x_2} \frac{d x}{\sqrt{1-x \, x_2}}\, \left(1-  \frac{1}{1-
  R^2 x}\right)\,\ln (R^2 x)\,.
\eq 
One can now split the integration domain $x_2 \in [0,1] \times x \in [0,1/x_2]$
as  $x \in [0,1] \times x_2 \in [0,1]$ and $  x \in [1,\infty ]
  \times x_2 \in [0,1/x ]\,.$
The integral corresponding to the second domain is identical to the first one
after exchanging $Q_1$ and $Q_2$.
Performing the integration over $x_2$ in the first domain, one  
gets for $K$
\bea
\label{K3}
&& K \,= \, \frac{1}{8\, 4^2 Q_1^2}\intfeyn  \frac{d x}{x^{5/2}} \,
  \left[ -6 \, \sqrt{x}\,(1\,+\,x) + (3\,+\,2\, x\,+\,3 x^2)(2\,  \ln
  (1+\sqrt{x}) \, \ln(1 \, - \, x))\right] \nonumber\\
&& \times \,\left(1-  \frac{1}{1-
  R^2 x}\right)\,\ln (R^2 x)\, + (1 \leftrightarrow 2)\,.
\eea
This one dimensional integral, after the change of variable $t =\sqrt{x}$  
can finally be reexpressed as
\bea
\label{K4}
&& K \,= \,- \frac{1}{64 \, Q_1 \, Q_2}\intfeyn  d t \left \{\frac{6 \, R}{t^2}(-2
\,t\,+\,\ln(1+t)-\ln(1-t))\,\ln(R \, t)\, -\,6 \,(1+R^2)\right.\nonumber \\
&&\left. \times \, \left(\frac{1}{1-R \, t}-\frac{1}{1+R \, t} \right)
 \ln (R \, t)  - \frac{6}{R} \ln (1+t)\, \ln \,(R \,t)
\,+\,\frac{6}{R} \ln (1-t)\, \ln \,(R \,t) \right.\nonumber \\
&&\left.+\,R \,\left (3 \, R^2 \, + \, 2 \, + \,\frac{3}{R^2}\right)
\left(\frac{ \ln(1+t) \, \ln (R \,t)}{1-R \, t}+
\frac{\ln(1+t) \, \ln (R \,t)}{1+R \, t}-\frac{\ln(1-t) \, \ln (R \,t)}{1-R \, t}
\right. \right.\nonumber \\
&&\left. \left. - \frac{\ln(1-t) \, \ln (R \,t)}{1+R \, t} \right) 
\right\} + \left(R \leftrightarrow \frac{1}{R}\right)\;.
\eea
The integrals arising from the two first lines of the previous expression,
supplemented
by the corresponding $R \to 1/R$ contribution, are easily computed by
integration by parts through logarithmic and polylogarithmic function 
${\rm Li}_2.$
Using Landen relation for ${\rm Li}_2$ (see Chap.1 of Ref.\cite{lewin}), this simplifies into 
\beq
\label{A}
 A=- \frac{1}{64 \, Q_1 \, Q_2} \left[6 \, \left(R + \frac{1}{R}\right) \ln^2
   R \, + \, 12\,
\left(R-\frac{1}{R}\right) \ln R 
%\nonumber \right.\\
%&&\hspace{-.8cm}\left.
+\, 12 \,  \left(R + \frac{1}{R}\right) \right]
\eq
The two last lines of Eq.(\ref{K4}) (denoted $B$ in the following) 
contain terms of the generic form
\beq
\label{generic}
\int \frac{\ln(a+ b x) \, \ln(c+e x)}{f+g x} g \, dx \;, 
\eq
which can be reduced to the standard form (see Chap.8 of Ref.\cite{lewin})
\beq
\label{Redgeneric}
\int \ln(1-y) \, \ln(1-c y) \frac{dy}{y}\;.
\eq
This last integral is evaluated in Chap.8 of Ref.\cite{lewin},
for a restricted domain in $c.$ An analytic continuation of such a result
for the whole complex domain in $c$ can also be performed \cite{wallon}. Let us
first define 
\beq
\label{defphi}
\varphi(\alpha) = \arg(e^{i \alpha})-\alpha 
\eq
for any
$real$ alpha. With this definition, $\frac{\phi(\alpha)}{2 \pi}$
is a winding number
which counts the number of turns one has to make around 0 in order to bring back
$\alpha$ to a value inside the interval $]-\pi,\pi].$ Then, the results for
the integral (\ref{Redgeneric}) reads, for $x$ real
\bea
\label{ResRedgeneric}
&& \int_0^x \ln(1-y) \, \ln(1-c y) \frac{dy}{y} \, = \, {\rm Li_3} \left(\frac{1-c \,
    x}{1-x} \right) + {\rm Li_3}\left(\frac{1}{c}\right) + {\rm Li_3}(1) -{\rm Li_3}(1-c \, x)
    \nonumber \\
&&  -{\rm Li_3}(1-x)-{\rm Li_3} \left( \frac{1-c \, x}{c(1-x)}\right)+\ln (1-x) \, {\rm Li_2}(1-c
    \,x)-\ln(1-c \, x) \, {\rm Li_2}(x) \nonumber \\
&& + \, (\ln(1-c \, x) -\ln(1-x)) \,
    {\rm Li_2}\left(\frac{1}{c} \right)\,+ \, \frac{\pi^2}{6} \ln(1-x) + \frac{1}{2} \ln c
    \, \ln^2(1-x) \nonumber \\
&& \,-\,i\,  \pi \,c_1 \, \ln^2 c + i\, 2 \, \pi \,  c_1 \, \ln
    c\, \ln(1-c \,x)-i\, 2 \, \pi \,  c_1 \, \ln
    c\, \ln(1- \,x)-4 \, \pi^2 c_2 \, c_3 \, \ln c \nonumber \\
&&   + \, i\, 2 \, \pi \,(c_1-c_4-c_5) \ln (1-x) \, \ln(1-c \, x) \,-\, 4 \, \pi^2 c_2 \, c_3 \,
    \ln(1-x) \, +\, 4 \, \pi^2 c_2 \, c_3 \, \ln(1-c\,x) \nonumber \\
&&  \,-\,\pi \,(c_1 -c_4
    -c_5)\, i\, \ln^2(1-c\,x)\,+\,i \, \pi \,(-c_1+c_5) \, \ln^2 \, (1-x) + i
    \, 4 \, \pi^3\, c_2 \, c_3\,,
\eea
where
\bea
\label{defcoef}
&& c_1 \,=\, \frac{1}{2 \pi} \, \varphi\,(\arg(c-1)-\arg(c)-\arg(1-x))\,, \quad c_2
\,= \, \frac{1}{2 \pi} \, \varphi\,(-\arg(c)) \nonumber \\
&&  c_3
\,= \, \frac{1}{2 \pi} \, \varphi \,(-\arg(1-x)) \,, \quad   c_4
\,= \, \frac{1}{2 \pi} \, \varphi\,(\arg(c)+\arg(x)) \nonumber \\
&&{\rm  and} \quad  c_5
\,= \, \frac{1}{2 \pi} \, \varphi\,(\arg(x)+\arg(c-1)-\arg(1-x))\,. 
\eea
$c_1,$ $c_2,$ $c_3$ and $c_4$ take either $0$ or $+1$ values while $c_5$ can
take $-1,$ $0$ or $1$ values. 
From this formula one can get an analytical form for $B$. The obtained result is very lenghtly
and contains a large bunch of $\ln,$ ${\rm Li}_2$ and ${\rm Li}_3$ of various rational
functions of $R.$
A first simplification occurs when using
 the Landen relation for ${\rm Li}_3$ 
\beq
\label{LandenLi3}
{\rm Li_3}(x)={\rm Li_3}\left(\frac{1}{x}\right)-\frac{\pi^2}{6}\ln(-x)-\frac{1}{6}\ln^3(-x)+\frac{i}{2}
\varphi\,(\arg(1-x)-\arg(-x)) \, \ln^2 x \,,
\eq
(see Chap.6 of Ref. \cite{lewin}), which effectively enables one to get
combinations of ${\rm Li_3}$ of the only arguments $R$ and $-R.$

Then, one reduces the terms containing ${\rm Li_2}$ as combinations of
${\rm Li_2}$ 
of the only arguments $R$ and $-R.$ This is possible after 
 using Euler relations
for ${\rm Li_2}$, 
which reads
\beq
\label{Euler1}
{\rm Li}_2(z)+{\rm Li}_2(1-z) = \frac{\pi^2}{6}-\ln z \, \ln(1-z)\;,
\eq
and
\beq
\label{Euler2}
{\rm Li}_2(z)+{\rm Li}_2\left(\frac{1}{z}\right) =
\frac{\pi^2}{3}-\frac{1}{2}\ln^2 z \,+i\,(\arg(z-1)-\arg(1-z))\ln z \;,
\eq
as well as Landen relation
\beq
\label{Landen}
{\rm Li}_2(z)+{\rm Li}_2\left(\frac{-z}{1-z}\right) =
-\frac{1}{2}\ln^2 (1-z) \,+i\,\varphi \,(-\arg(1-z))\,(\ln (1-z)-\ln z -i \, \pi) \,.
\eq
The most non trivial transformation is based on Hill
formula (which enables one to expand the double variable function
 ${\rm Li}_2(x\, y)$)
continuated in the whole complex plane (see \cite{wallon} for details),
\bea
\label{Hill}
&&\hspace{-.65cm}{\rm Li}_2(x \, y) \,=\,{\rm Li}_2(x) + {\rm Li}_2(y) - {\rm Li}_2\left[\frac{x(1 - y)}{1 - x y}\right] - {\rm Li}_2\left[\frac{y(1 - x)}{1 - x y}\right] - 
    \ln\left[\frac{1 - x}{1 - x y}\right] \ln\left[\frac{1 - y}{1 - x
    y}\right] \nonumber \\
&&\hspace{-.65cm} + \,
    i \,  \varphi\,(\arg(1 - y)  - \arg(1 - x y)) \, \left\{\ln\left[\frac{1 - x}{1 - x y}\right] - 
          \ln\left[\frac{y(1 - x)}{1 - x y}\right]\right\}\nonumber \\
&&\hspace{-.65cm} + \, 
    i \, \varphi\,(\arg(1 - x)  - \arg(1 - x y)) \left\{\ln \left[\frac{1 - y}{1 - x y}\right] - 
          \ln\left[\frac{x(1 - y)}{1 - x y}\right]\right\}\nonumber \\  
 &&\hspace{-.65cm}  + \,  \varphi\,(\arg(1 - y)  - \arg(1 - x y)) \, \varphi\,(\arg(1 - x)  - \arg(1 - x
    y))\,. 
\eea
After a long succession of painful simplifications, one finally gets
\bea
\label{B}
&&\hspace{-.8cm}  B\, = \, - \frac{1}{64 \, Q_1 \, Q_2} \,
 \left(3 \,R^2 +\, 2\,+\frac{3}{R^2}\right) \left(\ln(1-R)\,\ln^2 R
-\ln(R+1)\,\ln^2 R -2 \, {\rm Li_2}(-R)\, \ln R \nonumber \right. \\
&&\left. +\, 2 \,{\rm Li_2}(R)\, \ln R \,+\,2\, {\rm Li_3}(-R)\,-\,2\, {\rm Li_3}(R)\right)\,.
\eea
Using equations (\ref{A}) and  (\ref{B}) in order to express $K=A+B$
and using Eq.(\ref{MCalthreshold}), one proves the result (\ref{resultMmin}).

The limits $R \gg 1$ and $R \ll 1$ can be easily extracted using the
asymptotical formulae for large argument of ${\rm Li_2}$ and ${\rm Li_3.}$
In the case of ${\rm Li_2}$ they read
\beq
\label{Li2asymp}
{\rm Li_2}(x) \sim -\frac{1}{2} \ln^2 x +i (arg(x) -arg(-x)) \ln x + \frac{\pi^2}{3}
-\frac{1}{x} -\frac{1}{4 x^2} -\frac{1}{9 x^3} + {\rm o} \left(\frac{1}{x^3}\right)\;,
\eq
which reduces in the case of interest here to
\beq
\label{Li2asympP}
{\rm Li_2}(x) \sim -\frac{1}{2} \ln^2 x -i \, \pi \, \ln x + \frac{\pi^2}{3}
-\frac{1}{x} -\frac{1}{4 x^2} -\frac{1}{9 x^3} + {\rm o}
\left(\frac{1}{x^3}\right)\quad {\rm for} \quad x \to +\infty
\;,
\eq
\beq
\label{Li2asympM}
{\rm Li_2}(x) \sim -\frac{1}{2} \ln^2 x +i \, \pi \, \ln x + \frac{\pi^2}{3}
-\frac{1}{x} -\frac{1}{4 x^2} -\frac{1}{9 x^3} + {\rm o}
  \left(\frac{1}{x^3}\right) \quad {\rm for} \quad x \to -\infty \;.
\eq
Similarly for ${\rm Li_3}$ one gets from the functional relation (\ref{LandenLi3})
the asymptotic expansion
\beq
\label{Li3asymp}
{\rm Li_3}(x) \sim -\frac{1}{6}\ln^3 (-x) - \frac{\pi^2}{6} \ln(-x)\,+\, \frac{1}{x} + \frac{1}{8 x^2} +  \frac{1}{27 x^3} 
 + {\rm o} \left(\frac{1}{x^3}\right) \quad {\rm for} \quad x \to \infty \;,
\eq
which reduces to
\beq
\label{Li3asympP}
{\rm Li_3}(x) \sim -\frac{1}{6}\ln^3 x -i \frac{\pi}{2} \ln^2 x + \frac{\pi^2}{3}
\ln x + \frac{1}{x} + \frac{1}{8 x^2} +  \frac{1}{27 x^3} + {\rm o}
\left(\frac{1}{x^3}\right) \quad {\rm for} \quad x \to +\infty \;,
\eq
and 
\beq
\label{Li3asympM}
{\rm Li_3}(x) \sim -\frac{1}{6}\ln^3 x +i \frac{\pi}{2} \ln^2 x + \frac{\pi^2}{3}
\ln x + \frac{1}{x} + \frac{1}{8 x^2} +  \frac{1}{27 x^3} + {\rm o} \left(\frac{1}{x^3}\right) \quad {\rm for} \quad x \to -\infty\;.
\eq

These asymptotical formulae immediately lead
to 
\beq
\label{Mtminasymptotic}
{\cal M}_{t_{\min}} \sim +i s \frac{N_c^2-1}{N_c^2} \, \alpha_s^2 \, \alpha_{em} \,
\alpha(k_1) \, \beta(k_2) \, f_\rho^2 \, \frac{96 \pi^2}{Q_1^2\, Q_2^2}
\left(\frac{\ln R}{R}-\frac{1}{6 R} \right)\;.
\eq

%%%%%%%%%%%%%%%%%%%%%%%%%%%%%%%%%%%%%%%%%%%%%%%%

\subsection{Computation of the amplitude at $t=t_{min}$ in the partonic approach}
\label{tminparton}

In this appendix we rederive the asymptotic result
(\ref{Mtminasymptotic}) in a way which corresponds to the usual parton collinear limit.
In that limit, this simple results agrees with the fact that 
each loop of $t$-channel gluons gives rise to at most one logarithmic term.
The logarithmic term corresponds to the leading logarithm approximation (LLA), while the constant term goes beyond this approximation.
Let us show that this result can be easily obtained from the representation
(\ref{defPP}). Indeed, in the partonic approach, $k_\perp^2$ is
neglected
with respect to any scale of the order of $Q_1^2,$ and symmetrically any scale
of  the order 
of $Q_2^2$ is neglected with respect to $k_\perp^2.$
It is clear that in such an approximation, one immediately recovers the
dominant contribution $\frac{\ln R}{R}$ of (\ref{Mtminasymptotic}). The
question arises how to define a precise prescription in order to get the full
leading twist expression (\ref{Mtminasymptotic}).
It turns out that the integral (\ref{defK}), 
provides the proper result 
\beq
\label{limiteK}
K \,  \sim \, \frac{1}{3 Q_1 \, Q_2}\left(\frac{\ln R}{R}-\frac{1}{6 R} \right)\,,
\eq
after expanding the integrand
${\cal K}$ at
leading order, namely
\beq
\label{leadingIntegrand}
{\cal K} \sim \frac{1}{k^2 Q_1^2 z_1 \, \zb_1}\,,
\eq
and then integrating the result from
 $z_2 \, \zb_2 /R$ 
to $R \, z_1 \, \zb_1\,.$
Let us justify this prescription. We rewrite 
 the integral $K$ of Eq.(\ref{defK}) as
\beq
\label{defKu}
K= \frac{4}{Q_1 Q_2}\int\limits_0^{\frac{1}{2}} d z_1
\,\int\limits_0^{\frac{1}{2}} d z_2 \, \int\limits_0^\infty \, d u \frac{z_1
  \, \zb_1 z_2 \, \zb_2}{(u+R \, z_1 \, \zb_1)(u + z_2 \, \zb_2/R)} \,,
\eq 
and separate the $u =  k^2/(Q_1 Q_2)$ integration as
\beq
\label{sepKu}
 \int\limits_0^\infty \, d u \,=   \,\int\limits_{\beta R z_1 \zb_1}^{\infty}\, d u \, + \,
 \,\int\limits_{\alpha \frac{z_2 \zb_2}{R}}^{\beta R z_1 \zb_1}\, d u \,+\,\int\limits_0^{\alpha \frac{z_2 \zb_2}{R}} \, d u \,
\eq 
where the 
 parameters $\alpha$ and $\beta$ are arbitrary. In the large $R$ limit, $\beta
 R z_1 \zb_1 < \alpha \frac{z_2 \zb_2}{R}$ for $z_1 < \frac{\alpha}{\beta} \frac{z_2 \zb_2}{R^2}.$
Let us perform a systematic expansion of $K$ in the limit where 
 $\alpha$ and $\beta$ satisfy $R \gg \alpha \gg 1$ and  $R \gg 1/\beta \gg 1.$
 We organize the expansion in such a form that the large $R$ limit is taken
 first (which means the dominant twist approximation), and only then the large
 $\alpha$  and small $\beta$ limit are taken.
Decompose $K=K_1+K_2+K_3+K_4+K_5+K_6,$ where
\bea
\label{defK1}
&&K_1= \frac{4}{Q_1 Q_2}\int\limits_0^{\frac{1}{2}} d z_2
\,\int\limits_0^{\frac{\alpha}{\beta}\frac{z_2 \zb_2}{R^2}} d z_1 \, \int\limits_{\alpha
  \,\frac{z_2 \zb_2}{R}}^\infty \, d u \frac{z_1 \, \zb_1 z_2 \, \zb_2}{(u+R \,z_1
  \, \zb_1)(u + z_2 \, \zb_2/R)} \\
\label{defK2}
&&K_2= \frac{4}{Q_1 Q_2}\int\limits_0^{\frac{1}{2}} d z_2
\,\int\limits_0^{\frac{\alpha}{\beta}\frac{z_2 \zb_2}{R^2}} d z_1 \,
\int\limits_{\beta \, R \, z_1 \, \zb_1}^{\alpha \,\frac{z_2 \zb_2}{R}} \, d u
\frac{z_1 \, \zb_1 z_2 \, \zb_2}{(u+R \,z_1 \, \zb_1)(u + z_2 \, \zb_2/R)} \\
\label{defK3}
&&K_3= \frac{4}{Q_1 Q_2}\int\limits_0^{\frac{1}{2}} d z_2
\,\int\limits_0^{\frac{\alpha}{\beta}\frac{z_2 \zb_2}{R^2}} d z_1 \,
\int\limits_0^{\beta \, R \, z_1 \, \zb_1} \, d u \frac{z_1 \, \zb_1 z_2 \,
  \zb_2}{(u+R \,z_1 \, \zb_1)(u + z_2 \, \zb_2/R)} \\
\label{defK4}
&&K_4= \frac{4}{Q_1 Q_2}\int\limits_0^{\frac{1}{2}} d z_2
\,\int\limits_{\frac{\alpha}{\beta}\frac{z_2 \zb_2}{R^2}}^{\frac{1}{2}} d z_1 \,
\int\limits_{\beta \, R\,z_1 \zb_1}^\infty \, d u \frac{z_1 \, \zb_1 z_2 \,
  \zb_2}{(u+R \,z_1 \, \zb_1)(u + z_2 \, \zb_2/R)} \\
\label{defK5}
&&K_5= \frac{4}{Q_1 Q_2}\int\limits_0^{\frac{1}{2}} d z_2
\,\int\limits_{\frac{\alpha}{\beta}\frac{z_2 \zb_2}{R^2}}^{\frac{1}{2}} d z_1 \,
\int\limits_{\alpha \frac{\,z_2 \zb_2}{R}}^{\beta \, R \, z_1 \, \zb_1} \, d u
\frac{z_1 \, \zb_1 z_2 \, \zb_2}{(u+R \, z_1 \, \zb_1)(u + z_2 \, \zb_2/R)} \\
\label{defK6}
&&K_6= \frac{4}{Q_1 Q_2}\int\limits_0^{\frac{1}{2}} d z_2
\,\int\limits_{\frac{\alpha}{\beta}\frac{z_2 \zb_2}{R^2}}^{\frac{1}{2}} d z_1 \,
\int\limits_0^{\alpha \frac{\,z_2 \zb_2}{R}} \, d u \frac{z_1 \, \zb_1 z_2 \,
  \zb_2}{(u+R \, z_1 \, \zb_1)(u + z_2 \, \zb_2/R)} \,.
\eea
The integrals $K_1,$ $K_2$ and $K_3$ are of order $1/R^3$ and can thus be
neglected at leading twist. It corresponds to the absence of end-point
singularities in $z$ variables and  means that one could safely replace the lower
bound of $z_1$ integration in $K_4,$ $K_5$ and $K_6$ by 0.
 Let us focus now on the integral $K_5.$
The integration on $u$ runs from $\alpha \,z_2 \zb_2/R$ to
${\beta \, R \, z_1 \, \zb_1}.$ Since $z_1 \, \zb_1 \leq 1/4$ and 
$z_2 \, \zb_2 \leq 1/4,$ in the limit $R \gg 1,$
$\alpha \gg 1$ and  $\beta \ll 1,$ one can safely expand the integrand of
$K_5$ in powers of $u/(R z_1 \, \zb_1)$ and $z_2 \, \zb_2/(R \, u).$ 
At leading twist, only the dominant term has to be kept, and ${\cal K}$
can be approximated by  
(\ref{leadingIntegrand}). Integration with respect to $u,$ $z_1$ and $z_2$
then leads to
\beq
\label{limiteKalphabeta}
K_5 \,  \sim \, \left(\frac{\ln R}{R}-\frac{1}{6 R}\,+ \,\frac{1}{3 R}\ln \frac{\beta}{\alpha} \right) \,.
\eq
The logarithmic contribution $\ln \frac{\beta}{\alpha}$ corresponds to boundary
effect and is to be completely compensated by $K_4$ and $K_6,$ which behaves
respectively as $1/R \ln \beta$ and $1/R \ln \alpha$ in the limit $R \gg 1,$
$\alpha \gg 1$ and  $\beta \ll 1.$
This justify the assumption stated at the beginning of this paragraph.

%%%%%%%%%%%%%%%%%%%%%%%%%%%%%%%%%%%%%%%%%%%%%%%%

\subsection{Integrals}
\label{integrals}

In this appendix we collect all the generic integrals which appear
in the computation of the Born amplitude.
In addition to the integral  $I_2$ defined in Eq.(\ref{I2}), we introduce
the two following integrals with effective masses
 \beq
\label{I2m}
I_{2m}=\int \frac{d^d \kb}{\kb^2 ((\kb - \pb)^2+m^2)}\,,
\eq
and
\beq
\label{I2mm}
J_{2m_a m_b}=\int \frac{d^d \kb}{((\kb-\ab)^2+m_a^2) ((\kb-\bb)^2+m_b^2)}\,,
\eq

Let us first consider the case of the integral $I_{2m}$.
Using
Feynman parametrization, one easily get
\beq
\label{FeynI2m}
I_{2m}=\pi^{1+\epsilon} \Gamma(1-\epsilon) (\pb^2 + m^2)^{-1 + \eps} \intfeyn d \alpha
\alpha^{\eps-1} \left(1-\alpha \frac{\pb^2}{\pb^2 +m^2} \right )^{\eps-1}\;.
\eq
In the massless case, one immediately obtains
\beq
\label{resultI2}
I_{2}= \frac{2 \pi}{\pb^2 \eps}(1 + \eps (\ln (\pi \pb^2)-\Psi(1)))\;.
\eq
For the non zero mass case, the $\alpha$ integration leads to the hypergeometric function $_2F_1,$
\beq
\label{hypI2m}
I_{2m}=\pi^{1+\epsilon} \frac{\Gamma(1-\epsilon)}{\eps (\pb^2+m^2)^{1-\eps}} {}_2F_1\left(1-\eps,\eps,1+\eps;\frac{\pb^2}{\pb^2+m^2} \right)\;.
\eq
After performing the Euler transformation $z \to z/(z-1)$ in the argument of the hypergeometric function and then expanding the result in power of $\eps,$
one gets
\beq
\label{resultI2m}
I_{2m}=\frac{\pi}{\eps(\pb^2 +m^2)} (1 + \eps (\ln \pi - \Psi(1) - \ln
m^2+2 \ln(\pb^2+m^2)))\,.
\eq
Let us now turn to the more general case where the propagators contain two (different) masses. 
In this case dimensional regularization is not necessary, and after straightforward calculations, using Feynman parametrization, one obtains
\bea
\label{resultI2mm}
&&J_{2m_am_b}=\frac{\pi}{\sqrt{((\ab-\bb)^2 +(m_a-m_b)^2)((\ab-\bb)^2 +(m_a+m_b)^2)}} \\
&&\hspace{-.5cm} \times \ln \left |\frac{(\ab-\bb)^2 +m_a^2+m_b^2+\sqrt{((\ab-\bb)^2 +(m_a-m_b)^2)((\ab-\bb)^2 +(m_a+m_b)^2)}}{(\ab-\bb)^2 +m_a^2+m_b^2-\sqrt{((\ab-\bb)^2 +(m_a-m_b)^2)((\ab-\bb)^2 +(m_a+m_b)^2)}} \right |\,.\nonumber
\eea
For the purpose of our computation, we will need the previous integral
only for the special case where $\ab$ and $\bb$ are collinear.

In that case, one obtains, with $r=|\rb|,$
\bea
\label{tildeI2mm}
&&J_{2\alpha\beta}=\int \frac{d^d \kb}{((\kb-\rb a)^2+\alpha^2)
((\kb-\rb b)^2+\beta^2)}\nonumber \\
&&=\frac{\pi}{\sqrt{\lambda}}  \ln \frac{r^2(a-b)^2 +\alpha^2
+\beta^2 +\sqrt{\lambda}}{r^2(a-b)^2 +\alpha^2
+\beta^2 -\sqrt{\lambda}}\,,\nonumber
\eea
where we introduce the notation
\beq
\label{deffonctionlambda}
\lambda(x,y,z)=x^2+y^2+z^2-2 x y -2 x z -2 y z\,,
\eq
which  enables us to define, for the purpose of our computation,
\beq
\label{deflambda}
\lambda=\lambda(-r^2 (a-b)^2,\alpha^2,\beta^2)=(\alpha^2 -\beta^2)^2 +2(\alpha^2+\beta^2)r^2 (a-b)^2 + r^4
(a-b)^4\,.
\eq

%As for the $I_2$ integral, in addition to the integral  $I_{3m}$ 
%defined in Eq.(\ref{I3m}), we introduce
%the following integral with effective masses $m_a$ and $m_b$
%\beq
%\label{I3mm}
%I_{3mm}=\int \frac{d^d \kb}{\kb^2((\kb-\ab)^2+m_a^2) ((\kb-\bb)^2+m_b^2)}\,.
%\eq
Let us first consider the integral $I_{3m}.$
We start from the following identity
\bea
\label{identityI3m}
&&\int  \frac{d^2 \kb}{\kb^2(\kb-\pb)^2} \left( \frac{1}{(\kb-\ab)^2 + m^2}-
\frac{1}{\ab^2 + m^2} - \frac{1}{(\pb-\ab)^2 + m^2}+\frac{1}{(\kb-\pb+\ab)^2+m^2} \right) \nonumber \\
&& = -\left(\frac{1}{\ab^2 + m^2} + \frac{1}{(\pb-\ab)^2 + m^2} \right)\int \frac{d^d \kb}{\kb^2(\kb-\pb)^2}\nonumber \\
&& + \int  \frac{d^d \kb}{\kb^2(\kb-\pb)^2((\kb-\ab)^2 + m^2)}+  \int  \frac{d^d \kb}{\kb^2(\kb-\pb)^2((\kb-\pb+\ab)^2 + m^2)}\,.
\eea
This identity relates a finite expression on the lhs with a sum of 
dimensionally regularized integrals.
After shifting integration variable in the last integral on rhs of Eq.(\ref{identityI3m}), one obtains
\bea
\label{calculI3m}
&&\hspace{-1.2cm} \int  \frac{d^d \kb}{\kb^2(\kb-\pb)^2((\kb-\ab)^2 + m^2)} = \frac{1}{2} 
\left(\frac{1}{\ab^2 + m^2} + \frac{1}{(\pb-\ab)^2 + m^2} \right)\int \frac{d^d \kb}{\kb^2(\kb-\pb)^2}  \\
&& \hspace{-1.3cm} + \frac{1}{2} \int  \frac{d^2 \kb}{\kb^2(\kb-\pb)^2} \left( \frac{1}{(\kb-\ab)^2 + m^2}-
\frac{1}{\ab^2 + m^2} - \frac{1}{(\pb-\ab)^2 + m^2}+\frac{1}{(\kb-\pb+\ab)^2+m^2} \right),\nonumber
\eea 
which expresses $I_{3m}$ in terms of the divergent integral $I_2$ already
calculated and of a finite integral whose computation is our next task.
The method which we use is the generalization for the massive case, in momentum space, 
of the technique of calculation of 
massless two dimensional diagrams in the coordinate space encountered in conformal field theories \cite{vassiliev}.

The essential point is to perform in two dimensional finite integrals
 a conformal transformation of type
$\lb \to \lb/\lb^2$ on the integration variables and vector parameters, 
and of type $m^2 \to 1/m^2$ for the dimensionful parameters. This
transformation reduces the number of propagators.

Let us illustrate this method in the special case where $\ab$ and $\pb$
are collinear, which is of practical interest for our computation.
We will thus focus on the finite integral in rhs 
of Eq.(\ref{calculI3m}), which turns out to be $J_{3 m}$, as defined in
Eq.(\ref{defJ3m}).
The transformation 
\beq
\label{invtransf1}
\kb \to \frac{\Kb}{\Kb^2}, \quad \rb \to \frac{\Rb}{\Rb^2}, \quad m
 \to \frac{1}{M}
\eq
reduces the number of propagators and gives
 \bea
\label{2confJ3m}
&&J_{3 m} =\int  \frac{d^2 \kb}{\kb^2(\kb-\rb)^2} \left( \frac{1}{(\kb-\rb a)^2 + m^2}-
\frac{1}{\rb^2 + m^2} + (a \lr \aab) \right)\nonumber \\
&& = R^2  \int \frac{d^2 \Kb}{(\Kb-\Rb)^2} \left( \frac{ K^2
R^2}{(\Rb -a \Kb)^2 + \frac{\Kb^2 \Rb^2}{M^2}} - \frac{1}{a^2 r^2 + m^2}
+ (a \lr \aab) \right) \,.
\eea
After performing the shift of variable $\Kb= \Rb + \kb'$ and then 
finally making the inverse transformation
\beq
\label{invtransf2}
\kb' \to \frac{\kb}{\kb^2}, \quad \Rb \to \frac{\rb}{\rb^2}, \quad M
 \to \frac{1}{m}\,,
\eq
we end up with
 \beq
\label{2confJ3m2}
J_{3 m} =\frac{1}{r^2} \int  \frac{d^2 \kb}{\kb^2}  \left[\frac{(\rb +
\kb)^2}{(r^2 a^2 + m^2)(\left(\kb - \rb \frac{r^2 a \, \ab - m^2}{r^2 \ab^2 +
m^2}\right)^2 + \frac{m^2 r^4}{(r^2 \ab^2 + m^2)^2} )}-\frac{1}{a^2 r^2
+ m^2} + (a \lr \aab) \right]\,.
\eq
The computation of this integral can now be performed using standard
Feynman parameter technique.
It results in
\bea
\label{resultJ3m}
&&J_{3 m} = \frac{2 \pi}{r^2} \left\{ \left(\frac{1}{r^2 a^2 + m^2}-\frac{1}{r^2 \aab^2 + m^2}
\right) \ln \frac{r^2 a^2 +m^2}{r^2 \aab^2 + m^2} \right.\\
&&\left.+\left(\frac{1}{r^2 a^2
+ m^2}
+\frac{1}{r^2 \aab^2 + m^2}+ \frac{2}{r^2 a \aab - m^2}\right) \ln
\frac{(r^2 a^2 + m^2)(r^2 \aab^2 + m^2)}{m^2 r^2} \right\}\,.
\eea

Let us now focus on the $I_{4mm}$ integral.
We start with an analogous identity as those in Eq.(\ref{identityI3m}), which leads to
\bea
\label{calculI4mm}
&&\int  \frac{d^d \kb}{\kb^2(\kb-\pb)^2((\kb-\ab)^2 + m_a^2)(\kb-\bb)^2 + m_b^2)}  \\
&& \hspace{-1.3cm}= \frac{1}{2} 
\left(\frac{1}{(\ab^2 + m_a^2)(\bb^2+m_b^2)} + \frac{1}{((\pb-\ab)^2 + m_a^2)((\pb-\bb)^2 + m_b^2)} \right)\int \frac{d^d \kb}{\kb^2(\kb-\pb)^2} + \frac{1}{2} \, J\,,\nonumber
\eea 
where
\bea
\label{defI}
&& \hspace{-1.3cm} J =  \int  \frac{d^2 \kb}{\kb^2(\kb-\pb)^2} \left( \frac{1}{((\kb-\ab)^2 + m_a^2)((\kb-\bb)^2 + m_b^2)}-
\frac{1}{(\ab^2 + m_a^2)(\bb^2 + m_b^2)}\right. \\
&& \hspace{-1.3cm}\left. - \frac{1}{((\pb-\ab)^2 + m_a^2)((\pb-\bb)^2 + m_b^2)}+\frac{1}{((\kb-\pb+\ab)^2+m_a^2)((\kb-\pb+\bb)^2+m_b^2)} \right) \,.\nonumber
\eea 
This identity express the integral $I_{4m_am_b}$ as a sum of a IR divergent integral which has already been computed in Eq.(\ref{resultI2}) and of a finite integral $J$, on which we will apply 
the same trick based on conformal
 transformations. Once more, although our method can be applied to the previous
integral $J,$ we will restrict ourselves to the more simple case where $a$ and $b$ are collinear. It thus means
that we need to compute the integral $J_{4 \mu_1 \mu_2}$ as defined in Eq.(\ref{defJ4mm}).
We apply on $J_{4 \mu_1 \mu_2}$ three successive transformations accompanied by appropriate changes of variables as described above: conformal transformation, shift of integration variable and inverse conformal transformation. This gives
\bea
\label{calculJ4mm}
&&J_{4 \mu_1 \mu_2} = \frac{1}{\rb^2} \int \frac{d^d \kb}{\kb^2} \\
&&\hspace{-.7cm} \times \! \left \{\!\frac{\left((\kb-\rb)^2\right)^2}
 {\left(r^2 \zb_1+\mu_1^2\right)\!\left(\!\left(\kb+\rb \frac{z_1 \zb_1 r^2 - \mu_1^2}{\zb_1^2 r^2
  +\mu_1^2}\right)^2\!+\! \frac{r^4 \mu_1^2}{(r^2
z_1^2+\mu_1^2)^2}\right)\!\left(r^2 \zb_2+\mu_2^2\right)
\!\left(\!\left(\kb+\rb \frac{z_2 \zb_2 r^2 - \mu_2^2}{\zb_2^2 r^2
  +\mu_2^2}\right)^2\!+ \!\frac{r^4 \mu_2^2}{(r^2 z_2^2+\mu_2^2)^2}\right)}\right. \nonumber \\
&& \left.-\frac{1}{(r^2
 \zb_1+\mu_1^2)(r^2 \zb_2+\mu_2^2)}+ (z \lr \zb) \right \} \,. \nonumber 
\eea
In this expression, the integral which remains to be computed has the form
\beq
\label{defJ}
J_{3 \alpha \beta} =  \int \frac{d^2 \kb((\kb-\rb)^2)^2}{\kb^2 [(\kb-\rb
    a)^2+\alpha^2]\,[(\kb-\rb b)^2+\beta^2]}\,,
\eq
This integral can be rewritten in the form
\beq
\label{sepJ}
J_{3 \alpha \beta} =  J^{UV}_{3 \alpha \beta}+ J^{IR}_{3 \alpha \beta} 
\eq
where
\beq
\label{defJUV}
J^{UV}_{3 \alpha \beta}= \int \frac{d^d \kb \, \kb^2}{[(\kb-\rb
    a)^2+\alpha^2]\,[(\kb-\rb b)^2+\beta^2]}\,
\eq
and 
\beq
\label{defJIR}
J^{IR}_{3 \alpha \beta}=\int \frac{d^d \kb \, [((\kb-\rb)^2)^2-(\kb^2)^2]}{\kb^2 [(\kb-\rb
    a)^2+\alpha^2]\,[(\kb-\rb b)^2+\beta^2]}\,.
\eq
$J^{UV}_{3 \alpha \beta}$ is 
IR finite but UV divergent. On the contrary $J^{IR}_{3 \alpha \beta}$ is  UV finite but IR divergent.
The calculation of $J^{UV}_{3 \alpha \beta}$ is standard, although somehow lenghty, and leads to the result
\bea
\label{JUVresult}
&&J^{UV}_{3 \alpha \beta}=\pi \left \{ -\frac{\pi^\eps \Gamma(1-\eps)}{\eps}
- \frac{1}{2}\ln (\alpha^2 \, \beta^2) + \frac{\alpha^2-\beta^2}{2 r^2(a-b)^2} \ln \frac{\alpha^2}{\beta^2} \right.\\
&&\left. -\frac{\sqrt{\la}}{2 r^2 (a-b)^2} \ln \frac{r^2  (a-b)^2+\alpha^2
  +\beta^2+\sqrt{\la}}{r^2  (a-b)^2+\alpha^2
  +\beta^2-\sqrt{\la}}\right.\nonumber
\\
&&\left.-\frac{[\sqrt{\la}-\alpha^2+\beta^2-r^2(a^2-b^2)]^2}{4
r^2 (a-b)^2 \sqrt{\la}}\ln
\frac{\sqrt{\la}-r^2(a-b)^2-\alpha^2+\beta^2}{\sqrt{\la}+r^2(a-b)^2-\alpha^2+\beta^2}\right.\nonumber
\\
&&\left.-\frac{[\sqrt{\la}-\beta^2+\alpha^2-r^2(b^2-a^2)]^2}{4
r^2 (a-b)^2 \sqrt{\la}}\ln
\frac{\sqrt{\la}-r^2(a-b)^2+\alpha^2-\beta^2}{\sqrt{\la}+r^2(a-b)^2+\alpha^2-\beta^2}
\right\}\,.
\eea
The evaluation of  $J^{IR}_{3 \alpha \beta}$ can be simplified by expressing
it in the following way:
\bea
\label{decomposeJ3IR}
&&\!\!\!\!\! J^{IR}_{3 \alpha \beta}=\int d^d \kb \frac{2 \rb^2}{[(\kb-\rb
    a)^2+\alpha^2]\,[(\kb-\rb b)^2+\beta^2]}+r^4 \int \frac{d^d \kb}{\kb^2 [(\kb-\rb
    a)^2+\alpha^2]\,[(\kb-\rb b)^2+\beta^2]} \nonumber \\
&&-\,4 r^2 \int \frac{d^d \kb \, \kb
   \cdot \rb}{\kb^2 [(\kb-\rb
    a)^2+\alpha^2]\,[(\kb-\rb b)^2+\beta^2]}-4 \int \frac{d^d \kb \, \kb
   \cdot \rb}{[(\kb-\rb
    a)^2+\alpha^2]\,[(\kb-\rb b)^2+\beta^2]}\nonumber \\
&&+\,4 \int \frac{d^d \kb \, (\kb
   \cdot \rb)^2}{\kb^2 [(\kb-\rb
    a)^2+\alpha^2]\,[(\kb-\rb b)^2+\beta^2]}\,.
\eea
Each of these integrals is regular, except the second one, which diverges like 
\beq
\label{divJ3IR}
J^{IR\, div.}_{3 \alpha \beta} \equiv  \frac{\pi^{(1+\eps)}}{
 \eps \,  \Gamma(1+\eps)}\frac{r^4}{(r^2 a^2 +\alpha^2)(r^2 b^2 +\beta^2)}\,.
\eq
The evaluation of each of these integrals is lenghty but straighforward after
 using Feynman parametrization.

Since IR and UV divergences cancel when evaluating the integral I defined in 
Eq.(\ref{calculJ4mm}), we effectively need the finite part of $J_{3 \alpha
  \beta},$
defined as the remaining term after removing the pole in $1/\eps,$
which sums up the finite part of $J^{IR}_{3 \alpha \beta}$ and of  $J^{UV}_{3
  \alpha \beta}.$
With this definition, one can express $J_{4 \mu_1 \mu_2}$ as
\bea
\label{resI}
J_{4 \mu_1 \mu_2}= \hspace{-.2cm}\\
&&\hspace{-2.1cm} \frac{1}{r^2 (r^2 \zb_1^2 + \mu_1^2) (r^2 \zb_2^2 + \mu_2^2)}J^{finite}_{3
  \alpha \beta}\!\left(\frac{-z_1 \zb_1 r^2 +\mu_1^2}{\zb_1^2 r^2+\mu_1^2},
\frac{-z_2 \zb_2 r^2 +\mu_2^2}{\zb_2^2 r^2+\mu_2^2}, \frac{r^2 \mu_1}{r^2
  \zb_1^2 + \mu_1^2}, \frac{r^2 \mu_2}{r^2
  \zb_2^2 + \mu_2^2}, r \right) \nonumber \\
&&\hspace{-2.1cm}+ \frac{1}{r^2 (r^2 z_1^2 + \mu_1^2) (r^2 z_2^2 + \mu_2^2)}J^{finite}_{3
  \alpha \beta}\!\left(\frac{-z_1 \zb_1 r^2 +\mu_1^2}{z_1^2 r^2+\mu_1^2},
\frac{-z_2 \zb_2 r^2 +\mu_2^2}{z_2^2 r^2+\mu_2^2}, \frac{r^2 \mu_1}{r^2
  z_1^2 + \mu_1^2}, \frac{r^2 \mu_2}{r^2
  z_2^2 + \mu_2^2}, r \right) \nonumber
\eea
with
\bea
\label{resJ3finie}
&&J^{finite}_{3
  \alpha \beta}(a,\,b,\, \alpha,\, \beta,\, r)  \\
&&\hspace{-.8cm}=\pi \left\{\frac{1}{2 \rla}\left[-\frac{(\alpha-\beta)^2
  \,(\alpha+\beta)^2}{(a-b)^2 \,r^2} -4 \f{\alpha^2-\beta^2}{a-b}-2
  (\alpha^2+\beta^2)-((a-b)^2 +
  4(a+b)-12) r^2\right.\right.\nonumber \\
&&\hspace{-.8cm}\left. \left. -r^2 \left(\f{1}{r^2 a^2 + \alpha^2}+\f{1}{r^2 b^2 + \beta^2} \right) \right]\ln \frac{r^2  (a-b)^2+\alpha^2
  +\beta^2+\sqrt{\la}}{r^2  (a-b)^2+\alpha^2
  +\beta^2-\sqrt{\la}} \right.\nonumber \\
&&\hspace{-.8cm}\left.+\frac{1}{ \rla} \left[r^2 (a-b)^2  +2 (\alpha^2 + \beta^2) -2 \f{a\, \beta^2
  - b\, \alpha^2}{a-b} +2 a\, b \, r^2 + r^2 \f{\alpha^2 - \beta^2 +(a^2 -b^2)
  r^2}{b(r^2 a^2 +\alpha^2)-a(r^2 b^2 + \beta^2)} \right. \right. \nonumber \\
&&\hspace{-.8cm}\times \! \left. \left. \left(\! \f{r^2 a}{r^2 a^2 +
  \alpha^2}+ \f{r^2 b}{r^2 b^2 + \beta^2}-4\!\right)+2 \f{(\alpha^2-\beta^2)^2 +
  2 (\alpha^2+\beta^2)^2 \, r^2 (a-b)^2 + r^4 (a-b)^4}{b \, (r^2 a^2 +
  \alpha^2) -a \, (r^2 b^2 + \beta^2)}\f{1}{a-b} \right. \right. \nonumber \\
&&\hspace{-.8cm}\left. \left.+
  \f{(\alpha^2-\beta^2)^2}{r^2 (a-b)^2}\right]\ln
\frac{\sqrt{\la}+r^2(a-b)^2+\alpha^2-\beta^2}{\sqrt{\la}-r^2(a-b)^2+\alpha^2-\beta^2} +\left[ -\f{1}{a b (a-b)}.\f{a^2 \beta^2 -b^2 \alpha^2 + a b
  \rla}{a(r^2 b^2 + \beta^2)-b(r^2 a^2 +\alpha^2)}\right. \right.
  \nonumber \\
&&\hspace{-.8cm}\left. \left. -\f{r^2}{2 \rla}\left( \f{r^2 a}{r^2 a^2 +
  \alpha^2}+ \f{r^2 b}{r^2 b^2 + \beta^2}-4\right)\f{\rla + \alpha^2
  -\beta^2-r^2 (b^2 -a^2)}{a(r^2 b^2 + \beta^2)-b(r^2 a^2 +\alpha^2)}+
  \f{2}{a-b}+\f{a \alpha^2-b \beta^2}{(a-b) \rla}\right. \right. \nonumber \\
&&\hspace{-.8cm}\left. \left.+\f{(\alpha^2-\beta^2)^2}{2
  r^2 (a-b)^2 \rla}+\f{(a^2+b^2) r^2}{2
  \rla}-\f{a+b}{2(a-b)}\right]\ln\f{\alpha^2}{\beta^2}+\left[\f{1}{a b}-\f{r^4}{2(r^2 a^2 +\alpha^2)(r^2 b^2+\beta^2)}-\f{1}{2}
  \right]\right.\nonumber \\
&&\hspace{-.8cm}\times \left. \ln \f{\alpha^2 \,\beta^2}{r^4} + \f{1}{a(r^2 b^2 + \beta^2)-b(r^2 a^2 +\alpha^2)} \left[r^2 \left(\f{r^2 a}{r^2 a^2 +
  \alpha^2}+ \f{r^2 b}{r^2 b^2 + \beta^2}-4 \right)+\f{a^2 r^2 +
  \alpha^2}{a}\right. \right.\nonumber \\
&&\hspace{-.8cm}\left. \left.+\f{b^2 r^2 + \beta^2}{b} \right] \ln \f{r^2 a^2+ \alpha^2}{r^2
  b^2+ \beta^2}-\f{(r^2 a^2+ \alpha^2)(r^2 b^2 + \beta^2)-a b r^4}{a b (r^2 a^2+ \alpha^2)(r^2 b^2 + \beta^2)}\ln \f{(r^2 a^2+ \alpha^2)(r^2
  b^2+ \beta^2)}{r^4}\right\}\,,\nonumber
\eea
where appropriate additional $\ln r^2$ terms have been introduced after
  extracting the finite part of $J^{UV}_{3 \alpha \beta}$ and $J^{IR}_{3
  \alpha \beta}$
in order to write the final result
as made of logarithms of dimensionless quantities. This is possible since the
  final result $J_{4 \mu_1 \mu_2}$ is UV and IR finite.

\end{appendix}

%%%%%%%%%%%%%%%%%%%%%%%%%%%%%%%%%%%%%%%%%%%%%%%%%%%%%%%%%%%%%%%%%%%%%%%%%%%%

\end{document}